\documentclass[pre,preprint,superscriptaddress,amsmath,amssymb]{revtex4-1}
\usepackage{graphicx,subfigure,color}% Include figure files
\usepackage{dcolumn,amssymb,amsthm}% Align table columns on decimal point
\usepackage{bm,amsmath,longtable,booktabs, array,rotfloat}% bold math
 \usepackage{nomencl,multirow,longtable}

\begin{document}
%\preprint{}

\preprint{APS/123-QED}

\title{Discrete effect on the anti-bounce-back boundary condition of lattice Bhatnagar-Gross-Krook model for convection-diffusion equations}

\author{Liang Wang}
\affiliation{Key Laboratory of Condition Monitoring and Control for Power
Plant Equipment of Ministry of Eduction, North China Electric Power University,
Beijing 102206, China}
\affiliation{School of Energy Power and Mechanical Engineering, North China Electric Power University,
Beijing 102206, China}
\author{Xuhui Meng}
 \affiliation{Division of Applied Mathematics, Brown University, Providence, RI 02912, USA}
\author{Hao-Chi Wu}
 \affiliation{School of Control and Computer Engineering, North China Electric Power University, Beijing 102206, China}
\author{Tian-Hu Wang}
 \affiliation{Key Laboratory of Condition Monitoring and Control for Power
Plant Equipment of Ministry of Eduction, North China Electric Power University,
Beijing 102206, China}
\affiliation{School of Energy Power and Mechanical Engineering, North China Electric Power University,
Beijing 102206, China}
\author{Gui Lu}
 \email[Corresponding author:\quad]{lugui02@gmail.com}
 \affiliation{Key Laboratory of Condition Monitoring and Control for Power
Plant Equipment of Ministry of Eduction, North China Electric Power University,
Beijing 102206, China}
\affiliation{School of Energy Power and Mechanical Engineering, North China Electric Power University,
Beijing 102206, China}
%\author{Weifeng Zhao}
% \affiliation{Department of Applied Mathematics and Mechanics, University of Science and Technology Beijing, Beijing 100083, China}

\date{\today}% It is always \today, today,

\begin{abstract}
The discrete effect on the boundary condition has been a fundamental topic for the lattice Boltzmann method in simulating heat and mass transfer problems. In previous works based on the halfway anti-bounce-back (ABB) boundary condition for convection-diffusion equations (CDEs), it is reported that the discrete effect cannot be commonly removed in the Bhatnagar-Gross-Krook (BGK) model except for a special value of relaxation time. Targeting this point in the present paper, we still proceed within the framework of BGK model for two-dimensional CDEs, and analyze the discrete effect on a non-halfway ABB boundary condition which incorporates the effect of the distance ratio. By analyzing an unidirectional diffusion problem with a parabolic distribution, the theoretical derivations with three different discrete velocity models show that the numerical slip is a combined function of the relaxation time and the distance ratio. Different from previous works, we definitely find that the relaxation time can be freely adjusted by the distance ratio in a proper range to eliminate the numerical slip. Some numerical simulations are carried out to validate the theoretical derivations, and the numerical results for the cases of straight and curved boundaries confirm our theoretical analysis. Finally, it should be noted that the present analysis can be extended from the BGK model to other lattice Boltzmann (LB) collision models for CDEs, which can broaden the parameter range of the relaxation time to approach 0.5.

\end{abstract}

\maketitle

%\tableofcontents

\section{\label{intro}Introduction}
In the past couples of decades, the lattice Boltzmann method (LBM) has been gradually developed as an effective and powerful technique for a wide range of application areas \cite{Succi01,Guob13}, such as single-phase flows, multiphase flows, microgaseous flows and porous flows \cite{ChenSY98,Aidun10,HuangHB15,ZhangJ11,Liang19}. Unlike the conventional computational methods, the LBM solves the discrete Boltzmann equation instead of the macroscopic continuum equations. The kinetic nature of the LBM possesses several attractive features in flow simulations, such as simple program, intrinsically parallel computation, and easy boundary treatment. Among the other successful extensions, the LBM has also been adapted to solve convection-diffusion equations (CDEs), which are commonly encountered in studying heat and mass transfer associated with fluid flows. So far, there have been many LB models proposed for CDEs \cite{Sman00,Huang14,WangL18}. More detailed reviews about these works can be found in Refs. \cite{ShiB09,Yoshida10,Chai13,ZhangSub}.

To completely solve CDEs by the LBM, apart from the numerical algorithm for the LB equation (LBE), the boundary condition should also be specified for the unknown distribution functions at boundary nodes (i.e., lattice nodes nearest to the physical boundary). It is a critical issue and has attracted increasing researchers' efforts towards accurate boundary treatments. In several recent publications \cite{ZhangT12,ChenQ13,HuangJ15,Kruger17}, the reader can trace some existing LBM boundary conditions such as the ABB scheme \cite{Ginzburg05,Li17} and the non-equilibrium extrapolation scheme \cite{Chai16,GuoZ02}. The terminology of ABB is in contrast to the bounce-back (BB) scheme for fluid flows. That is, the outgoing population reflects back in the opposite direction with the BB scheme, while it changes its sign with the ABB scheme \cite{Ginzburg17}. As have recognized in the boundary conditions of LBM for flow simulations, it is known that the discrete effect on the boundary condition also must be minimized to derive correct results for CDEs \cite{ZhangT12,Dubois10,ShuC16}. However, there has not been extensive investigations on this topic as those for the fluid flow simulations. Based on the developed Taylor expansion method \cite{Dubois07,Dubois08}, Dubois \emph{et al}. \cite{Dubois10} analyzed the ABB boundary condition within the framework of multiple-relaxation-time (MRT) model for one-dimensional diffusion equation with the Dirichlet boundary condition. They demonstrated that the halfway ABB (HABB) boundary condition can be accurate up to order two in space under a specific combination of the relaxation rates. Within the BGK model framework, Zhang \emph{et al}. \cite{ZhangT12} proposed a HABB boundary condition for CDEs, and also analyzed the discrete effect of their boundary condition. For the diffusion in Couette flow with wall injection, they derived mathematically that the concentration jump or the numerical slip is related with the relaxation time and has a second-order dependence with the lattice spacing. It is also shown that the numerical slip cannot commonly be removed in the BGK model. As for the discrete effect of the HABB boundary condition, Cui \emph{et al.} \cite{ShuC16} revisit this topic based on the MRT model with three discrete lattice models, and derive the numerical slip relating with two relaxation rates and the square of lattice spacing. Their theoretical analysis and numerical results show that the discrete effect on the HABB boundary condition can be removed owing to the free relaxation parameter $s_2$ in the MRT model, while it cannot be eliminated except for a special value of the relaxation time in the BGK model. However, we note that the boundary condition in the above works is concentrated to the halfway boundary scheme, which intrinsically disregards the possible degree of freedom from the wall arrangements between lattice nodes.

Actually, in the boundary conditions for CDEs, the wall can be located between two lattice nodes with an arbitrary but not only halfway intersection distance \cite{Li17,HuangJ15,ChenQ13,Dubois19}. This means that if the wall location is embodied in the boundary condition, it may appear as a free parameter besides the relaxation time in the derived numerical slip. Therefore, it naturally brings out a fundamental question about the discrete effect of the boundary condition: whether the numerical slip can be eliminated while not limited at a special relaxation time in the BGK model. To our knowledge, no publications have been reported on this topic. In this work, we will analyze the discrete effect on the non-halfway ABB (NHABB) boundary condition for CDEs within the framework of BGK model. The boundary condition proposed in Ref. \cite{HuangJ15} is adopted here for its locality and ability to adjust the wall location arbitrarily between lattice nodes. And importantly, we will show how to choose the relaxation time to eliminate the discrete effect freely by tuning the parameter of wall location. From this point, in addition to resorting to other LBE models (e.g., MRT model) for more degrees of freedom, the present work reveals another way to eliminate the discrete effect on boundary condition of the BGK model for CDEs.

The paper is organized as follows. In Sec. \ref{Sec2}, the BGK-LBE for the CDE with a source term is presented. Sec. \ref{Sec3} is devoted to analyzing the discrete effect of the halfway and non-halfway ABB boundary conditions. In Sec. \ref{Sec4}, some numerical experiments and discussions are given, and followed by some conclusions finally presented in Sec. \ref{results}.

\section{Lattice Bhatnagar-Gross-Krook model for convection-diffusion equations}\label{Sec2}
In this work, our analyses are specially focused on the BGK model for the convection-diffusion equation. For the two-dimensional case, the CDE with a source term reads
\begin{equation}\label{CDES}
  \partial_t\phi+\nabla\cdot(\phi\bm{u})=\nabla\cdot(D\nabla\phi)+R,
\end{equation}
where $\phi$ is the scalar variable as a function of time and space, $D$ is the diffusion coefficient, $\bm{u}=(u_x,u_y)^T$ is the convection velocity with $T$ denoting the transposition operator, and $R(\bm{x}, t)$ is the source term.
The BGK-LBE to sovle the CDE \eqref{CDES} is written as follows
\begin{equation}\label{LBECDE}
  f_i(\bm{x}+\bm{c}_i\delta_t, t+\delta_t)-f_i(\bm{x}, t)=-\frac{1}{\tau_\phi}\left[f_i(\bm{x}, t)-f_i^{(eq)}(\bm{x}, t)\right]+\delta_t(1-\frac{1}{2 \tau_\phi})R_i(\bm{x}, t),
\end{equation}
where $\{f_i(\bm{x}, t): i=0,1,\cdots, b-1\}$ are the distribution functions associated with the discrete velocities $\{\bm{c}_i: i=0,1,\cdots, b-1\}$ at position $\bm{x}$ and time $t$, $\tau_\phi$ is the relaxation time, $\delta_t$ is the evolution time increment; $f_i^{(eq)}(\bm{x}, t)$ is the equilibrium distribution function, and $R_i(\bm{x}, t)$ is the discrete source term, which are respectively defined as
\begin{align}
   f_i^{(eq)}(\bm{x}, t)&=\omega_i\phi\left(1+\frac{\bm{c}_i\cdot\bm{u}}{c_s^2}\right),\\
   R_i(\bm{x}, t)&=\omega_i R,
\end{align}
where $\omega_i$ is the weight coefficient, and $c_s$ is the sound speed.

The discrete velocity set $\bm{c}_i$ is subjected to the $\text{D}n\text{Q}b$ ($\text{D}n\text{Q}b$ denotes $b$ velocity directions in $n$D space) lattice models reported in the literature \cite{Qian92}. In this work, the discrete effect of the ABB boundary condition is inspected with three discrete lattice models. As adopted in Ref. \cite{ShuC16} for the MRT collision model, the D2Q4, D2Q5 and D2Q9 models are also considered for subsequent analysis connected with the BGK model. The corresponding parameters for these three models are given as follows:
for the D2Q4 model, $\{\bm{c}_i: i=1,2,3,4\}=\{(\pm1,0)c,(0,\pm1)c\}$, $\omega_{1-4}=\frac{1}{4}$, and $c_s^2=\frac{1}{2}c^2$; for the D2Q5 model, $\{\bm{c}_i: i=0,1,2,3,4\}=\{(0,0)c, (\pm1,0)c,(0,\pm1)c\}$, $\omega_{0-4}=\frac{1}{5}$, and $c_s^2=\frac{2}{5}c^2$; for the D2Q9 model, $\{\bm{c}_i: i=0,1,\cdots,8\}=\{(0,0)c, (\pm1,0)c,(0,\pm1)c,(\pm1,\pm1)c\}$, $\omega_0=\frac{4}{9}$, $\omega_{1-4}=\frac{1}{9}$, $\omega_{5-8}=\frac{1}{36}$, and $c_s^2=\frac{1}{3}c^2$; where $c=\delta_x/\delta_t$ is the lattice speed with $\delta_x$ the lattice spacing.

The macroscopic variable $\phi$ is determined by the distribution functions as
\begin{equation}
   \phi(\bm{x}, t)=\sum_i f_i(\bm{x}, t)+\frac{\delta_t}{2}R(\bm{x}, t).
\end{equation}
With this definition, the CDE with a source term, Eq. \eqref{CDES}, can be recovered from the BGK model through the Chapman-Enskog analysis \cite{WangL18,Chai13}. Also, the diffusion coefficient can be derived and determined by the relaxation time $\tau_\phi$ as $D=c_s^2(\tau_\phi-\frac{1}{2})\delta_t$.

Numerically, the evolution of the BGK-LBE \eqref{LBECDE} is implemented via two steps, i.e., the collision and streaming step:
\begin{align}
   &\text{Collision}:\quad f_i^{*}(\bm{x},t)=f_i(\bm{x},t)-\frac{1}{\tau_\phi}\left[f_i(\bm{x}, t)-f_i^{(eq)}(\bm{x}, t)\right]+\delta_t(1-\frac{1}{2 \tau_\phi})R_i(\bm{x}, t),\notag\\
   &\text{Streaming}:\quad f_i(\bm{x}+\bm{c}_i\delta_t, t+\delta_t)=f_i^{*}(\bm{x},t),
  \end{align}
where $f_i^{*}(\bm{x},t)$ is the postcollision distribution function.

\section{Discrete effect on the anti-bounce-back boundary condition of BGK model for CDE}\label{Sec3}
\begin{figure}
\centering
\includegraphics[width=0.5\textwidth]{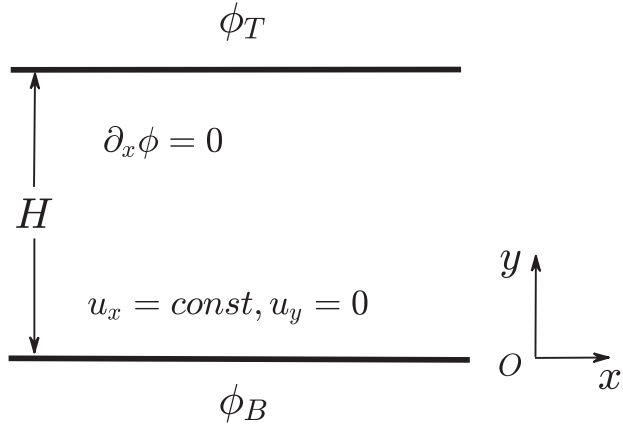}
\caption{Schematic of the unidirectional and time-independent diffusion problem.}
\label{ScheDiff}
\end{figure}

As mentioned previously, the discrete effect on the ABB boundary condition for CDEs is concentrated on the halfway scheme in the existing analysis \cite{ZhangT12,ShuC16}. However, the discrete effect is unclear when it is affected by the wall location between lattice nodes. To resolve this gap, we will restrict within the framework of BGK model to analyze the discrete effect of the NHABB boundary condition. For clarity of illustration, our analysis is based on the problem used in Ref. \cite{ShuC16} within the MRT framework. The considered problem is an unidirectional and time-independent diffusion in a straight channel (see Fig. \ref{ScheDiff}) in which $u_x$ is constant, $u_y=0$, and $\partial_x\phi=0$ for any scalar variable $\phi$.
For the constant $\phi_B$ and $\phi_T$ corresponding to the bottom and top walls (i.e., the Dirichlet boundary condition), the problem can be described by the following equations
\begin{subequations}\label{CDEeqs}
 \begin{equation}\label{SEqCDE}
    D \frac{\partial^2 \phi}{\partial y^2} + R =0,
  \end{equation}
  \begin{equation} \label{BoundCDE}
      \phi(x, y=0)=\phi_B, \quad  \phi(x, y=H)=\phi_T,
  \end{equation}
\end{subequations}
where $H$ is the height of the channel. As the source term $R$ is further defined by
\begin{equation}
  R=2D \Delta\phi/H^2, \quad \Delta\phi=\phi_T-\phi_B,
\end{equation}
we can obtain the analytical solution to this simple problem
\begin{equation}\label{Eq:Analytical}
  \phi(y)=\phi_B+\Delta\phi \frac{y}{H}(2-\frac{y}{H}).
\end{equation}

\begin{figure}
\centering
\includegraphics[width=0.65\textwidth]{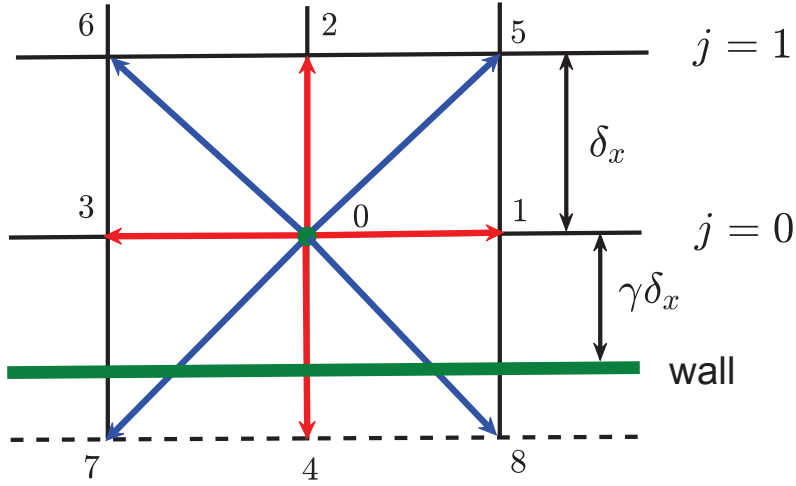}
\caption{Schematic of the boundary arrangement with an arbitrary distance in the D2Q$b$ ($b=4,5$, or $9$) lattice model. The bottom wall in the HABB boundary condition is placed with $\gamma=1/2$, while this restriction for the bottom wall is removed in the present analysis.}
\label{Schelattice}
\end{figure}

When the BGK model \eqref{LBECDE} is implemented to solve the above diffusion problem, after a time step $\delta_t$, the unknown distribution functions (see Fig. \ref{Schelattice}) should be specified by proper boundary conditions of the LBM.
To this end, we adopt the NHABB boundary condition proposed by Huang and Yong \cite{HuangJ15} using the asymptotic analysis technique. As displayed in Fig. \ref{Schelattice}, the wall boundary, say the bottom wall, is located away from its nearest inner lattice nodes with the distance of $\gamma\delta_x$. When the location of bottom wall is adjustable, the distance ratio $\gamma$ can be considered as a free parameter, which is conventionally used in the range of $0\leq\gamma\leq1$ \cite{ZhangSub,ChenQ13,HuangJ15}. However, as will be shown later, we would like to note that the distance ratio $\gamma$ would not be intuitively limited in $0\leq\gamma\leq1$, but can be larger than unity to derive accurate results. With this point, the unknown distribution functions at the layer $j=0$ are  then determined by the following equations.

D2Q4 or D2Q5 lattice model:
\begin{equation}\label{antiBoun45}
   f_2=\left(1-\frac{1}{2\gamma}\right)f_2^{*}-\frac{1}{2\gamma}f_4^{*}+\frac{\omega_2}{\gamma}\phi_B,
\end{equation}

D2Q9 lattice model:
\begin{subequations}\label{antiBoun9}
 \begin{equation}
    f_2=\left(1-\frac{1}{2\gamma}\right)f_2^{*}-\frac{1}{2\gamma}f_4^{*}+\frac{\omega_2}{\gamma}\phi_B,
  \end{equation}
  \begin{equation}
     f_5=\left(1-\frac{1}{2\gamma}\right)f_5^{*}-\frac{1}{2\gamma}f_7^{*}+\frac{\omega_5}{\gamma}\phi_B,
  \end{equation}
  \begin{equation}
     f_6=\left(1-\frac{1}{2\gamma}\right)f_6^{*}-\frac{1}{2\gamma}f_8^{*}+\frac{\omega_6}{\gamma}\phi_B.
  \end{equation}
\end{subequations}
One can see that if the parameter $\gamma=\frac{1}{2}$, this boundary condition will reduce to the HABB scheme \cite{ZhangT12,ShuC16}. It should be noted that the above boundary condition is a local scheme and has second-order accuracy for the case of straight walls \cite{HuangJ15}. Additionally, we would like to point out that following the procedures presented in Ref. \cite{Zhao17b} for Dirichlet boundary condition of the Navier-Stokes equations, the above boundary condition can also be obtained by the Maxwell iteration method \cite{Yong16,Zhao17a} with the diffusive scaling $\delta_t=\eta\delta_x^2$ and an adjustable parameter $\eta$.

Based on the adopted boundary schemes and the assumptions for the diffusion problem, one can follow the derivations in Refs. \cite{ZhangT12,ShuC16} to derive that
\begin{equation}\label{phiD2Q4}
  \phi_1=\frac{\gamma+1}{\gamma}\phi_0-\frac{\phi_B}{\gamma}-\frac{4\tau_\phi^2+8(\gamma-1)\tau_\phi+3}{2\gamma(2\tau_\phi-1)}\delta_t R
\end{equation}
for the D2Q4 lattice model, and
\begin{equation}\label{phiD2Q5}
  \phi_1=\frac{\gamma+1}{\gamma}\phi_0-\frac{\phi_B}{\gamma}-\frac{6\tau_\phi^2+(10\gamma-11)\tau_\phi+4}{2\gamma(2\tau_\phi-1)}\delta_t R
\end{equation}
for the D2Q5 lattice model, and
\begin{equation}\label{phiD2Q9}
  \phi_1=\frac{\gamma+1}{\gamma}\phi_0-\frac{\phi_B}{\gamma}-\frac{8\tau_\phi^2+2(6\gamma-7)\tau_\phi+5}{2\gamma(2\tau_\phi-1)}\delta_t R
\end{equation}
for the D2Q9 lattice model. Here, $\phi_0$ and $\phi_1$ are the scalar variables at the layer of $j=0$ and $j=1$.

During the above derivations, we can also deduce that the numerical scalar variable $\phi_j$ satisfies $D(\phi_{j+1}-2\phi_j+\phi_{j-1})=-R\delta_x^2$, where the diffusion coefficient $D$ is also given by $D=c_s^2(\tau_\phi-\frac{1}{2})\delta_t$. Clearly, this is the central finite-difference discretization of Eq. \eqref{SEqCDE}, meaning that the BGK-LBE is an equivalent solver for the CDEs. However, due to the discrete effect from the boundary condition, the LB results will deviate from the analytical solution to the problem [Eq. \eqref{Eq:Analytical}]. As a result, the solution of the BGK model with the NHABB boundary condition can be expressed as
\begin{equation}\label{phiNAna}
  \phi_j=\phi_B+\Delta\phi\frac{y_j}{H}\left(2-\frac{y_j}{H}\right)+\phi_s,
\end{equation}
where $y_j=(j+\gamma)\delta_x$, and $\phi_s$ is the numerical slip originated from the discrete effect of the boundary condition. By substituting $\phi_0$ and $\phi_1$ from Eq. \eqref{phiNAna} respectively into Eqs. \eqref{phiD2Q4}, \eqref{phiD2Q5} and \eqref{phiD2Q9}, we can obtain the numerical slips $\phi_s$ from the D2Q4, D2Q5 and D2Q9 lattice models.

D2Q4 lattice model:
\begin{subequations}\label{phisTD2Q4}
\begin{equation}\label{phisD2Q4}
  \phi_s=\frac{\Delta\phi}{4}\frac{\delta_x^2}{H^2}\left[4\tau_\phi^2+4\gamma(2\tau_\phi-1)-8\tau_\phi-4\gamma^2+3\right],
\end{equation}
  \begin{equation}\label{phishalfD2Q4}
   \gamma=\frac{1}{2} \Rightarrow \phi_s=\frac{\Delta\phi}{4}\frac{\delta_x^2}{H^2}\left(4\tau_\phi^2-4\tau_\phi\right).
  \end{equation}
\end{subequations}

D2Q5 lattice model:
\begin{subequations}\label{phisTD2Q5}
\begin{equation}\label{phisD2Q5}
  \phi_s=\frac{\Delta\phi}{5}\frac{\delta_x^2}{H^2}\left[6\tau_\phi^2+5\gamma(2\tau_\phi-1)-11\tau_\phi-5\gamma^2+4\right],
\end{equation}
 \begin{equation}\label{phishalfD2Q5}
   \gamma=\frac{1}{2} \Rightarrow \phi_s=\frac{\Delta\phi}{5}\frac{\delta_x^2}{H^2}\left(6\tau_\phi^2-6\tau_\phi+\frac{1}{4}\right).
  \end{equation}
\end{subequations}

D2Q9 lattice model:
\begin{subequations}\label{phisTD2Q9}
\begin{equation}\label{phisD2Q9}
  \phi_s=\frac{\Delta\phi}{6}\frac{\delta_x^2}{H^2}\left[8\tau_\phi^2+6\gamma(2\tau_\phi-1)-14\tau_\phi-6\gamma^2+5\right],
\end{equation}
\begin{equation}\label{phishalfD2Q9}
   \gamma=\frac{1}{2} \Rightarrow \phi_s=\frac{\Delta\phi}{6}\frac{\delta_x^2}{H^2}\left(8\tau_\phi^2-8\tau_\phi+\frac{1}{2}\right).
\end{equation}
\end{subequations}

From each of the above equations, one can find that the HABB and NHABB boundary conditions generate a nonzero numerical slip $\phi_s$, which has second-order accuracy in space owing to the term of $\delta_x^2/H^2$. It is noted that the results of $\phi_s$ for the HABB boundary condition ($\gamma=1/2$) [Eqs. \eqref{phishalfD2Q4}, \eqref{phishalfD2Q5} and \eqref{phishalfD2Q9}] here are identical to those of the BGK model given in Ref. \cite{ShuC16}. However, the numerical slip $\phi_s$ of the halfway boundary condition is not available for the non-halfway boundary condition. Due to the fixed location of wall with $\gamma=1/2$, $\phi_s$ of the HABB boundary condition is only related with $\tau_\phi$. This indicates that the discrete effect of the HABB boundary condition always exists in the BGK model unless a special relaxation time is used \cite{ZhangT12,ShuC16}. In contrast, owing to the adjustable distance ratio $\gamma$ as revealed above, $\phi_s$ of the NHABB boundary condition [Eqs. \eqref{antiBoun45} and \eqref{antiBoun9}] is dependent with the relaxation time $\tau_\phi$ and the distance ratio $\gamma$. Thus, the relaxation time $\tau_\phi$ has more degree of freedom to minimize the discrete effect on the boundary condition. The above results inspire us that the numerical slip $\phi_s$ of the BGK model could be eliminated freely by the relaxation time $\tau_\phi$ with the help of the free parameter $\gamma$.

Now let us focus on how to choose the relaxation time $\tau_\phi$ tuned by the distance ratio $\gamma$ to guarantee $\phi_s=0$. Mathematically, this can be done by solving the quadratic equation $\phi_s=0$ from Eqs. \eqref{phisD2Q4}, \eqref{phisD2Q5} and \eqref{phisD2Q9} respectively for the D2Q4, D2Q5 and D2Q9 discrete lattice model. Because of the stability condition as well as the positivity of diffusivity, there is only one root of $\tau_\phi$ that is determined by $\gamma$ from Eqs. \eqref{gamapsD2Q4}, \eqref{gamapsD2Q5} and \eqref{gamapsD2Q9}. For the HABB boundary scheme ($\gamma=1/2$), the corresponding relaxation time $\tau_\phi$ is obtained by Eqs. \eqref{gamapshalfD2Q4}, \eqref{gamapshalfD2Q5} and \eqref{gamapshalfD2Q9}.

D2Q4 lattice model:
\begin{subequations}\label{gamapsTD2Q4}
\begin{equation}\label{gamapsD2Q4}
  \phi_s=0\Rightarrow \tau_\phi=\frac{2(1-\gamma)+\sqrt{8\gamma^2-4\gamma+1}}{2},
\end{equation}
\begin{equation}\label{gamapshalfD2Q4}
   \gamma=\frac{1}{2}, ~ \phi_s=0 \Rightarrow  \tau_\phi=1.
\end{equation}
\end{subequations}

D2Q5 lattice model:
\begin{subequations}\label{gamapsTD2Q5}
\begin{equation}\label{gamapsD2Q5}
  \phi_s=0\Rightarrow \tau_\phi=\frac{11-10\gamma+\sqrt{5(44\gamma^2-20\gamma+5)}}{12},
\end{equation}
\begin{equation}\label{gamapshalfD2Q5}
   \gamma=\frac{1}{2}, ~ \phi_s=0 \Rightarrow  \tau_\phi=\frac{6+\sqrt{30}}{12}.
\end{equation}
\end{subequations}

\begin{figure}
\centering
\includegraphics[width=0.75\textwidth,height=0.65\textheight]{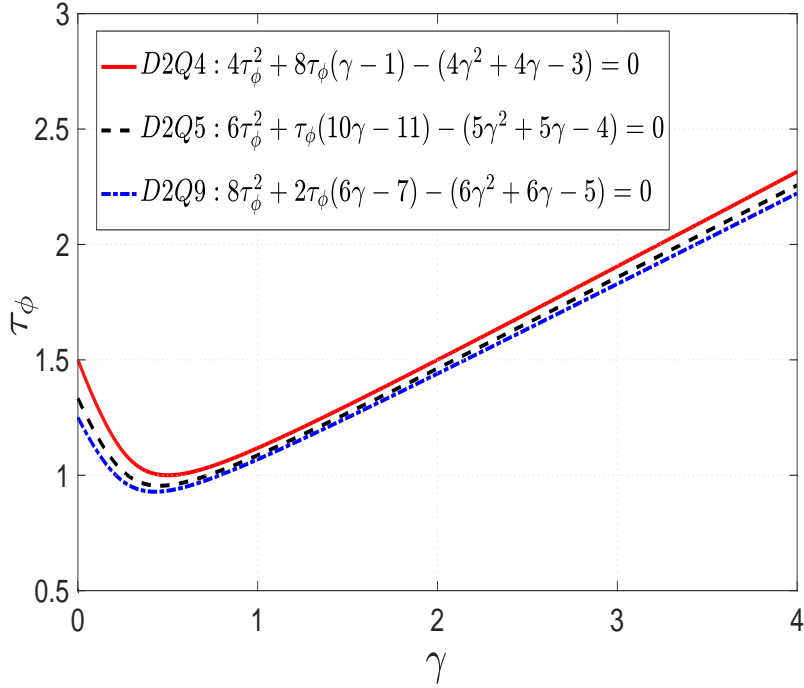}
\vspace{-8em}\caption{Dependence of $\tau_\phi$ on $\gamma$ resulted from $\phi_s=0$: $4\tau_\phi^2+8\tau_\phi(\gamma-1)-(4\gamma^2+4\gamma-3)=0$ for the D2Q4 lattice model; $6\tau_\phi^2+\tau_\phi(10\gamma-11)-(5\gamma^2+5\gamma-4)=0$ for the D2Q5 lattice model; $8\tau_\phi^2+2\tau_\phi(6\gamma-7)-(6\gamma^2+6\gamma-5)=0$ for the D2Q9 lattice model.}
\label{Depgatau}
\end{figure}

D2Q9 lattice model:
\begin{subequations}\label{gamapsTD2Q9}
\begin{equation}\label{gamapsD2Q9}
  \phi_s=0\Rightarrow \tau_\phi=\frac{7-6\gamma+\sqrt{3(28\gamma^2-12\gamma+3)}}{8},
\end{equation}
\begin{equation}\label{gamapshalfD2Q9}
   \gamma=\frac{1}{2}, ~ \phi_s=0 \Rightarrow  \tau_\phi=\frac{2+\sqrt{3}}{4}.
\end{equation}
\end{subequations}
For the case of $\gamma=1/2$, the corresponding values of $\tau_\phi$ are fixed and identical to those reported in Ref. \cite{ShuC16}. However, with the adopted NHABB boundary condition, the relaxation time $\tau_\phi$ is related with the distance ratio $\gamma$, and hence can be freely tuned by $\gamma$ to ensure $\phi_s=0$. To see this more clearly, the dependence of $\tau_\phi$ on $\gamma$ as given above for $\phi_s=0$ is shown in Fig \ref{Depgatau}.
Take the D2Q4 model as an example. It is seen that as the distance ratio $\gamma$ increases, $\tau_\phi$ can change continuously to fulfil $\phi_s=0$, while as $\gamma=1/2$ for the HABB boundary condition, $\phi_s=0$ determines the merely fixed $\tau_\phi=1$. When the distance ratio $\gamma$ varies in the region of $0\leq\gamma\leq1$, the relaxation time $\tau_\phi$ will take values limitedly between 1 and 1.5. To achieve a wider parameter range for $\tau_\phi$, the distance ratio $\gamma$ should not be confined to the region of $0\leq\gamma\leq1$. This point will be examined in the subsequent numerical examples, and the computations therein reveal that reasonable results can be also obtained as $\gamma$ is beyond 1. From the figure, it is also find that there are two values of $\gamma$ with $\gamma\leq2$ corresponding to the same $\tau_\phi$ in the range of $1\leq\tau_\phi\leq1.5$, while there is only one $\gamma$ with $\gamma>2$ corresponding to a certain $\tau_\phi>1.5$. Similar results stored in Eqs. \eqref{gamapsD2Q5} and \eqref{gamapsD2Q9} are also observed for the D2Q5 and D2Q9 lattice models. It should be noted that as $\gamma \rightarrow 0$, the boundary conditions \eqref{antiBoun45} and \eqref{antiBoun9} may lose the numerical stability since the included term $\frac{1}{\gamma}$ will become very large. This will be also affirmed in the subsequent numerical examples. Therefore, the relaxation time $\tau_\phi$ should be chosen carefully to avoid very small values of $\gamma$ in the computations. However, we would note that such limitation of parameter range in $\gamma$ may be remedied through recomposing the distribution functions and their coefficients in the adopted boundary condition \cite{ZhangSub}.

From the above derivations, it is clear that due to the distance ratio $\gamma$, the numerical slip $\phi_s$ can be theoretically eliminated within the framework of BGK model. As for the MRT model, it has been commonly recognized that the numerical slip $\phi_s$ can be overcome owing to its multiple relaxation parameters \cite{ShuC16}. For an explicit comparison of the two model frameworks, Table \ref{Tab:compareBGKMRT} presents the numerical slip $\phi_s$ and the relaxation parameter corresponding to $\phi_s=0$ derived in this work together with those deduced with the MRT model in Ref. \cite{ShuC16}.
\begin{table}[!hbp]
  \caption{Numerical slip $\phi_s$ and relaxation parameter corresponding to $\phi_s=0$ in the present work within the BGK model and those \cite{ShuC16} within the MRT model. The listed results are based on the D2Q4, D2Q5 and D2Q9 lattice models. The relaxation parameter $s_1=1/\tau_\phi$ is related to the diffusion coefficient in  \cite{ShuC16}, while $s_2$ is served as the free relaxation parameter.}
  \vspace{0.4em}
  \label{Tab:compareBGKMRT}
  \centering
  \begin{tabular*}
  {16cm}{@{\extracolsep{\fill}}ccc}
 \toprule[0.04em]\toprule[0.04em]
 \multirow{2}{*}{Discrete lattice model} &  \multicolumn{2}{c}{$\phi_s$}\\
 \cline{2-3}
 &    BGK(Present)&  MRT(Ref. \cite{ShuC16}) \\
 \midrule[0.04em]
 D2Q4&   $\Delta\phi\frac{\delta_x^2}{H^2}\frac{4\tau_\phi^2+4\gamma(2\tau_\phi-1)-8\tau_\phi-4\gamma^2+3}{4}$& $\Delta\phi\frac{\delta_x^2}{H^2}\frac{2-s_1-s_2}{2s_1s_2}$ \\
 D2Q5&   $\Delta\phi\frac{\delta_x^2}{H^2}\frac{6\tau_\phi^2+5\gamma(2\tau_\phi-1)-11\tau_\phi-5\gamma^2+4}{5}$&    $\Delta\phi\frac{\delta_x^2}{H^2}\frac{s_1s_2-12(s_1+s_2)+24}{20s_1s_2}$ \\
 D2Q9&   $\Delta\phi\frac{\delta_x^2}{H^2}\frac{8\tau_\phi^2+6\gamma(2\tau_\phi-1)-14\tau_\phi-6\gamma^2+5}{6}$&    $\Delta\phi\frac{\delta_x^2}{H^2}\frac{s_1s_2-8(s_1+s_2)+16}{12s_1s_2}$\\
  \midrule[0.05em]
  \multirow{2}{*}{Discrete lattice model} &  \multicolumn{2}{c}{Relaxation parameter($\phi_s=0$)} \\
 \cline{2-3}
&    BGK(Present)&  MRT(Ref. \cite{ShuC16}) \\
 \midrule[0.04em]
 D2Q4&  $\tau_\phi=\frac{2(1-\gamma)+\sqrt{8\gamma^2-4\gamma+1}}{2}$&  $s_2=2-s_1$ \\
 D2Q5&  $\tau_\phi=\frac{11-10\gamma+\sqrt{5(44\gamma^2-20\gamma+5)}}{12}$&  $s_2=\frac{12(s_1-2)}{s_1-12}$ \\
 D2Q9&  $\tau_\phi=\frac{7-6\gamma+\sqrt{3(28\gamma^2-12\gamma+3)}}{8}$&  $s_2=\frac{8(s_1-2)}{s_1-8}$ \\
 \bottomrule[0.04em]
 \bottomrule[0.04em]
\end{tabular*}
\end{table}
One can find that the relaxation parameter under $\phi_s=0$ is related with another relaxation rate in the MRT model, while it is related with the distance ratio $\gamma$ here in the BGK model. Based on this, we note that the elimination of numerical slip in the MRT model is ascribed to the degree of freedom from the relaxation parameter of the evolution equation, while in the BGK model here, the degree of freedom is from the wall location of the NHABB boundary condition.

\section{Numerical results and discussions}\label{Sec4}
To examine the above theoretical analysis, the BGK-LBE with the halfway and non-halfway ABB boundary conditions \cite{HuangJ15} are executed in the numerical simulations. The diffusion problems considered here are the same as those adopted in Ref. \cite{ShuC16}. In the following simulations, the lattice spacing $\delta_x$ is determined by $\delta_x= H/(M+2\gamma)$ with $\gamma$ the distance ratio, where $M$ is the grid number between the walls in the vertical direction. The distance ratio $\gamma$ is set as an input variable, and other related parameters are given by
\begin{equation}
  \delta_t=\eta \delta_x^2, \quad \eta=\frac{\chi(\tau_\phi-\frac{1}{2})}{D},
\end{equation}
where $\chi$ is a model-dependent constant defined by $c_s^2=\chi c^2$, and equals to $1/2, 2/5, 1/3$ respectively for the D2Q4, D2Q5 and D2Q9 lattice model.

\subsection{Unidirectional diffusion in a straight channel}
\begin{figure}
\centering
\includegraphics[width=0.75\textwidth]{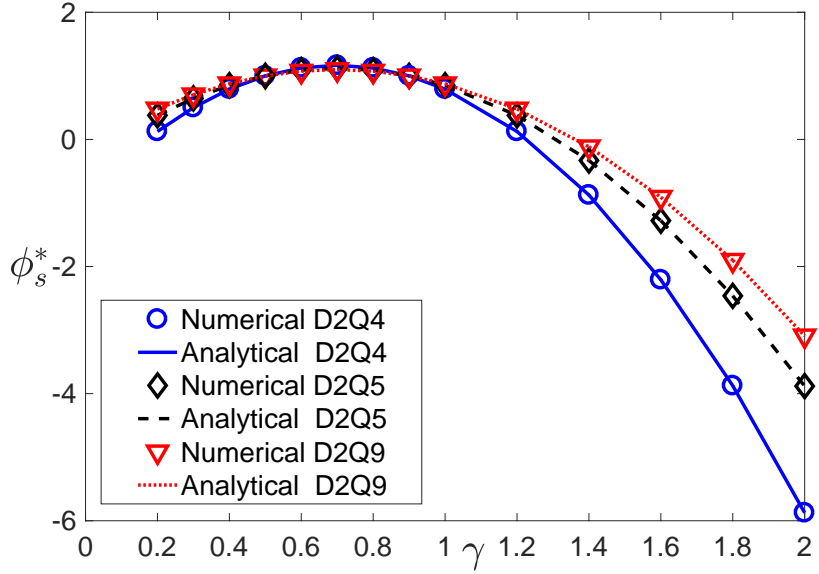}
\caption{ Normalized results of the numerical slip, $\phi_s^*=\phi_s/\phi_{s,\gamma=\frac{1}{2}}$, as a function of $\gamma$ at $\tau_\phi=1.2$ and $M=15$. $\phi_s$ denotes the numerical slip derived theoretically [Eqs. \eqref{phisD2Q4}, \eqref{phisD2Q5} and \eqref{phisD2Q9}] or predicted from numerical simulations. $\phi_{s,\gamma=\frac{1}{2}}$ denotes the numerical slip $\phi_s$ at the case of $\gamma=\frac{1}{2}$.}
\label{normalslip}
\end{figure}

The first problem is shown in Fig. \ref{ScheDiff}, where $H=1$, $u_x=0.1$, and the diffusion coefficient $D=0.1$. The periodic boundary condition is applied to the inlet and outlet of the channel, and the ABB boundary condition with the distance ratio $\gamma$ is applied to the top and bottom walls. In Fig. \ref{normalslip}, the simulated results of numerical slip, normalized by the results at the case of $\gamma=1/2$, are presented as a function of $\gamma$ at $\tau_\phi=1.2$ and $M=15$. The normalized theoretical results [Eqs. \eqref{phisD2Q4}, \eqref{phisD2Q5} and \eqref{phisD2Q9}] are also included for comparison. Clearly, the numerical predictions are well consistent with the theoretical derivations for the three discrete lattice models. In particular, the unambiguous agreement between such two results is observed when $\gamma$ is greater than $1$ up to $2$, and even at $\gamma=3$ (the results are not shown here).
Moreover, we find that the computations will break down as $\gamma$ decreases to 0.1. These twofold results verify the aforementioned statements about the choice of distance ratio $\gamma$ in the boundary conditions. Additionally, as the distance ratio $\gamma$ increases, it is observed that the numerical slip varies increasingly first and then decreasingly after one certain $\gamma$ due to its quadratic function as derived above. It is noted that similar results as shown in Fig. \ref{normalslip} can also be obtained at other relaxation times.
%and more conspicuously, its value changes from positive to negative, implying that the adopted boundary condition overestimates followed by underestimating the scalar fields.

\begin{figure}
\begin{tabular}{cc}
\includegraphics[width=0.55\textwidth,height=0.3\textheight]{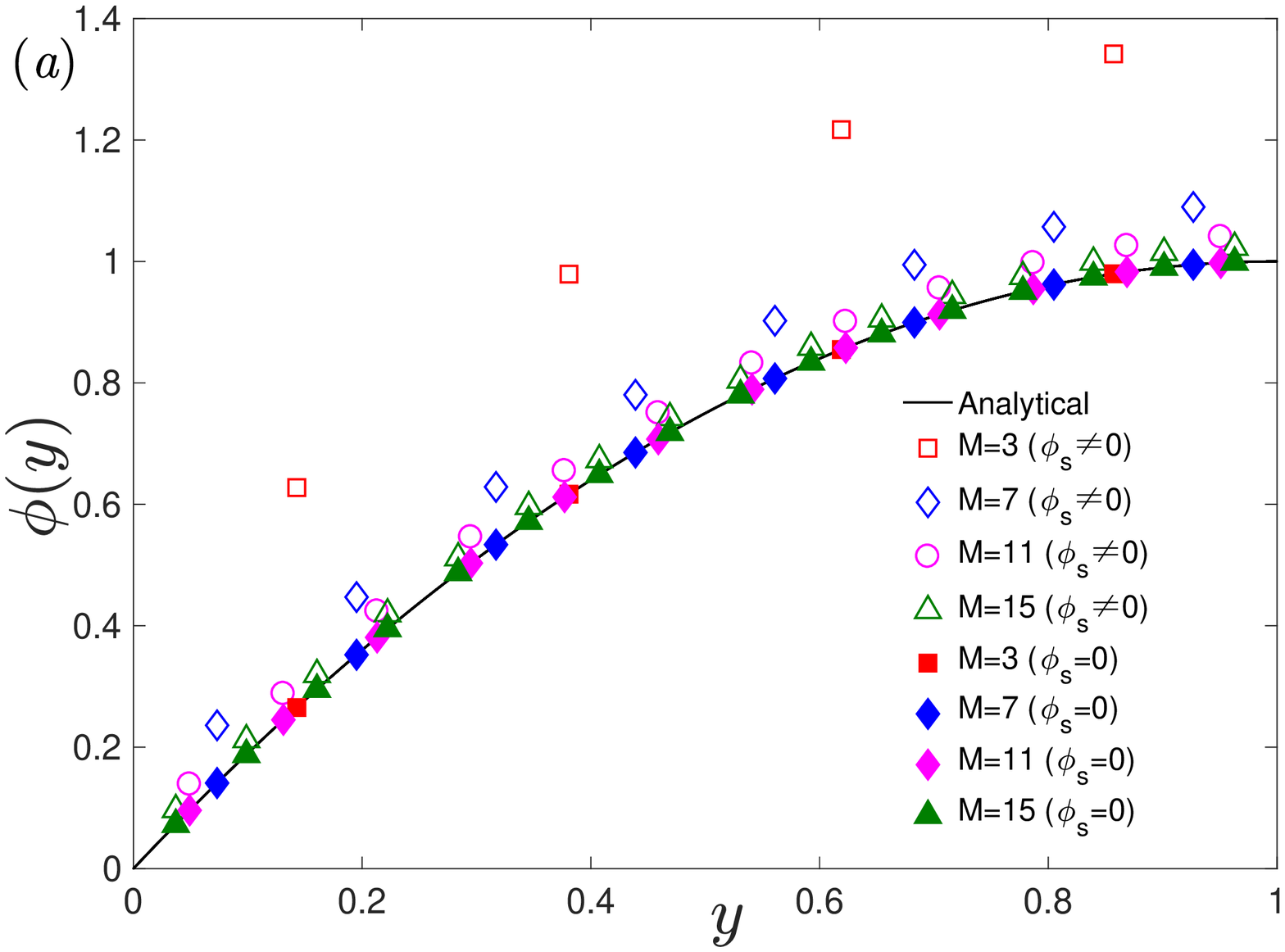}
\includegraphics[width=0.55\textwidth,height=0.3\textheight]{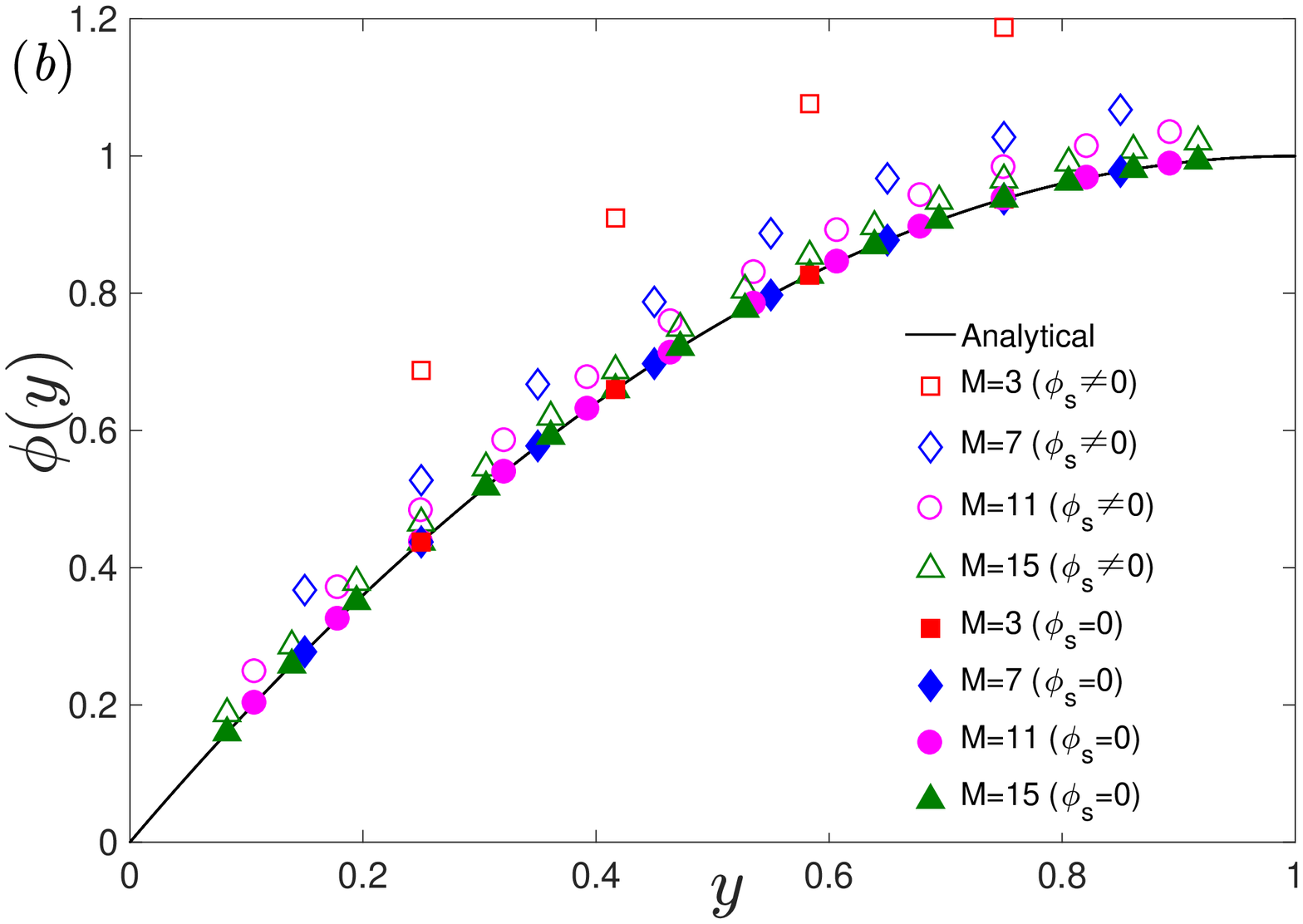}
%(a)&(b)\\
\end{tabular}
\caption{Profiles of scalar variable $\phi$ under different lattice sizes and [(a) $\gamma=0.6$; (b) $\gamma=1.5$] from the D2Q4 lattice model. Empty shapes denote the case that the relaxation time ($\tau_\phi=3.0$) dissatisfies $\phi_s=0$ with the distance ratio $\gamma$, while filled ones denote the case that $\tau_\phi$   satisfies $\phi_s=0$ with $\gamma$.}
\label{numSlipD2Q4}
\end{figure}

\begin{figure}
\begin{tabular}{cc}
\includegraphics[width=0.55\textwidth,height=0.3\textheight]{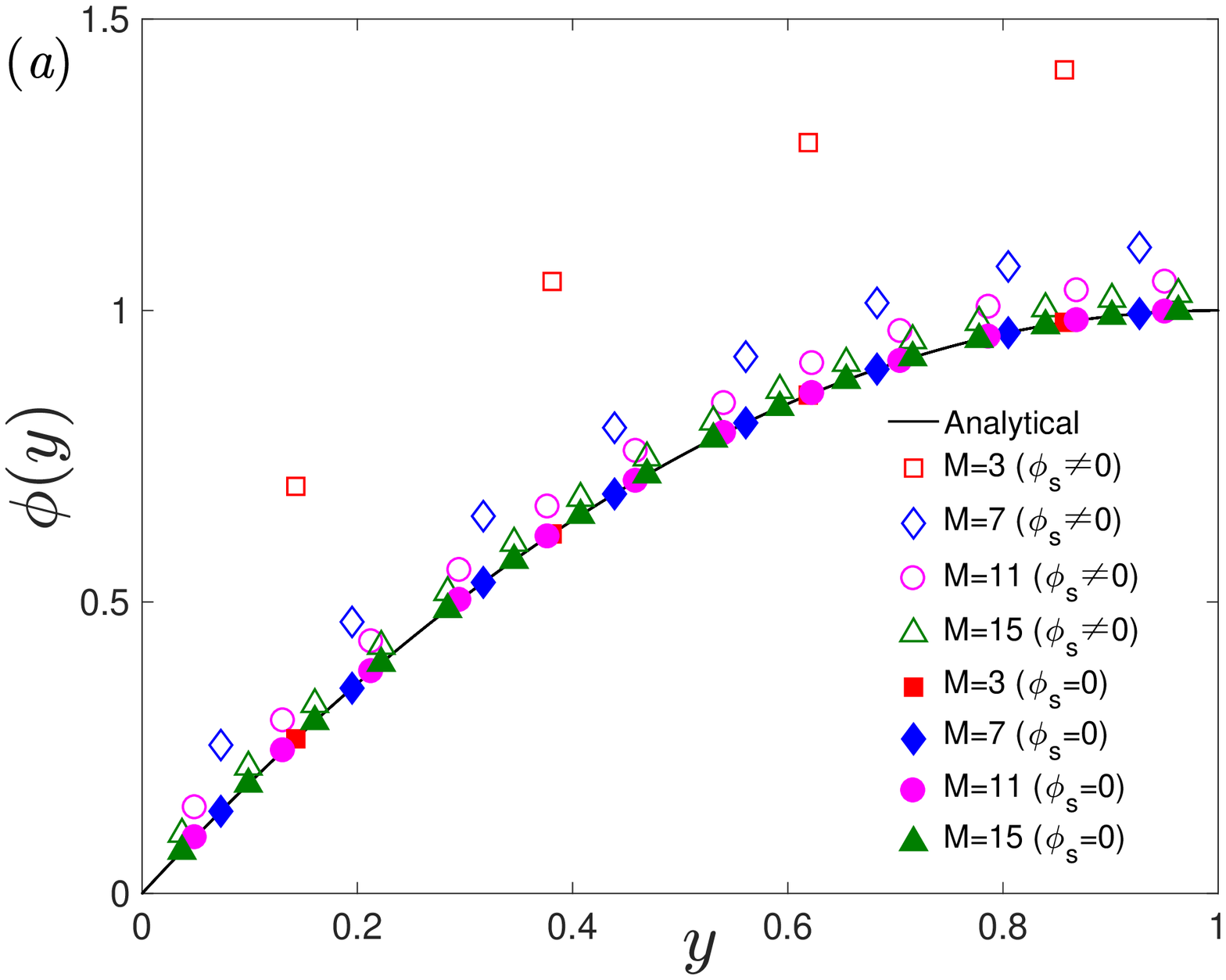}
\includegraphics[width=0.55\textwidth,height=0.3\textheight]{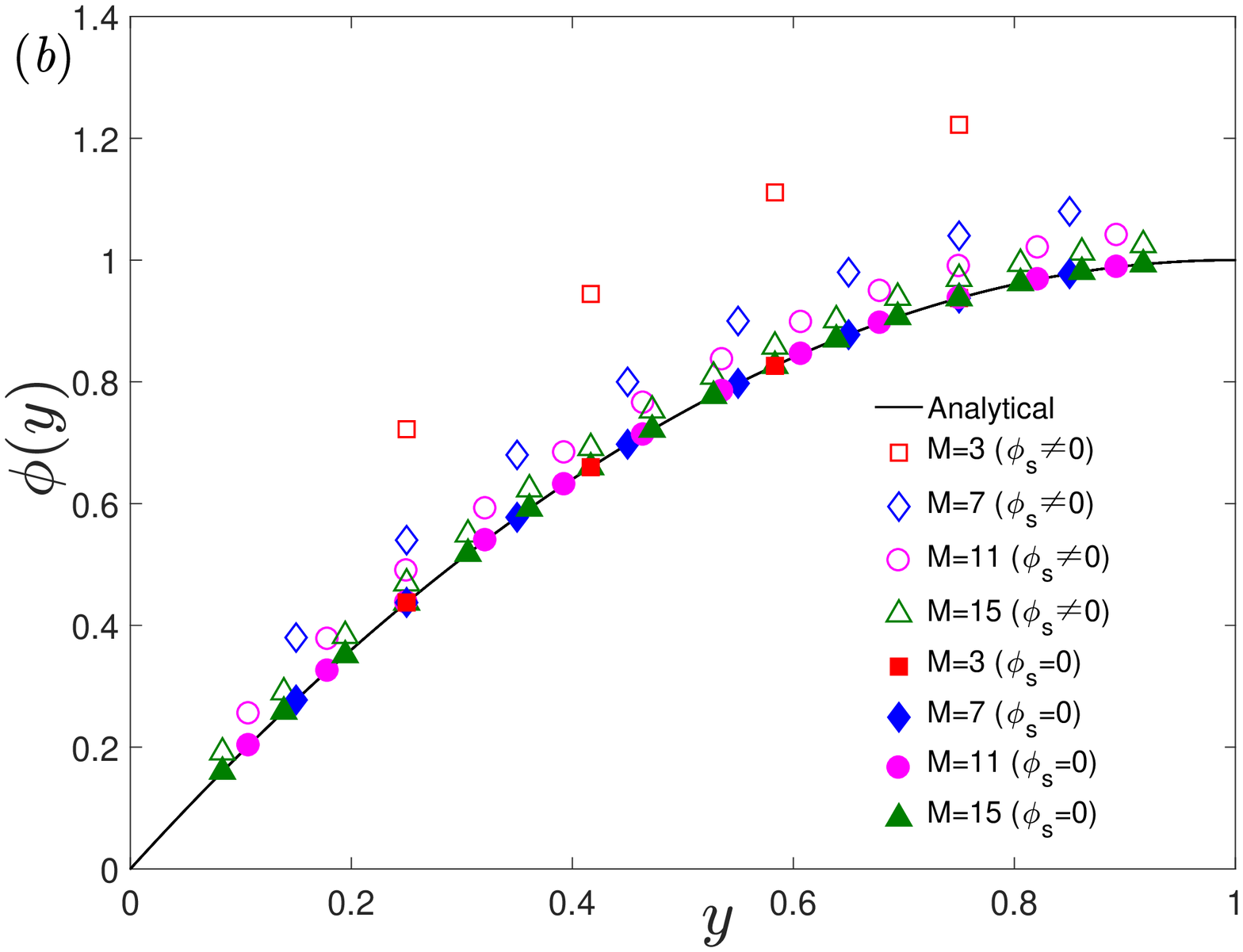}
%(a)&(b)\\
\end{tabular}
\caption{Profiles of scalar variable $\phi$ under different lattice sizes and [(a) $\gamma=0.6$; (b) $\gamma=1.5$] from the D2Q5 lattice model. Empty shapes denote the case that the relaxation time ($\tau_\phi=3.0$) dissatisfies $\phi_s=0$ with the distance ratio $\gamma$, while filled ones denote the case that $\tau_\phi$   satisfies $\phi_s=0$ with $\gamma$.}
\label{numSlipD2Q5}
\end{figure}

\begin{figure}
\begin{tabular}{cc}
\includegraphics[width=0.55\textwidth,height=0.3\textheight]{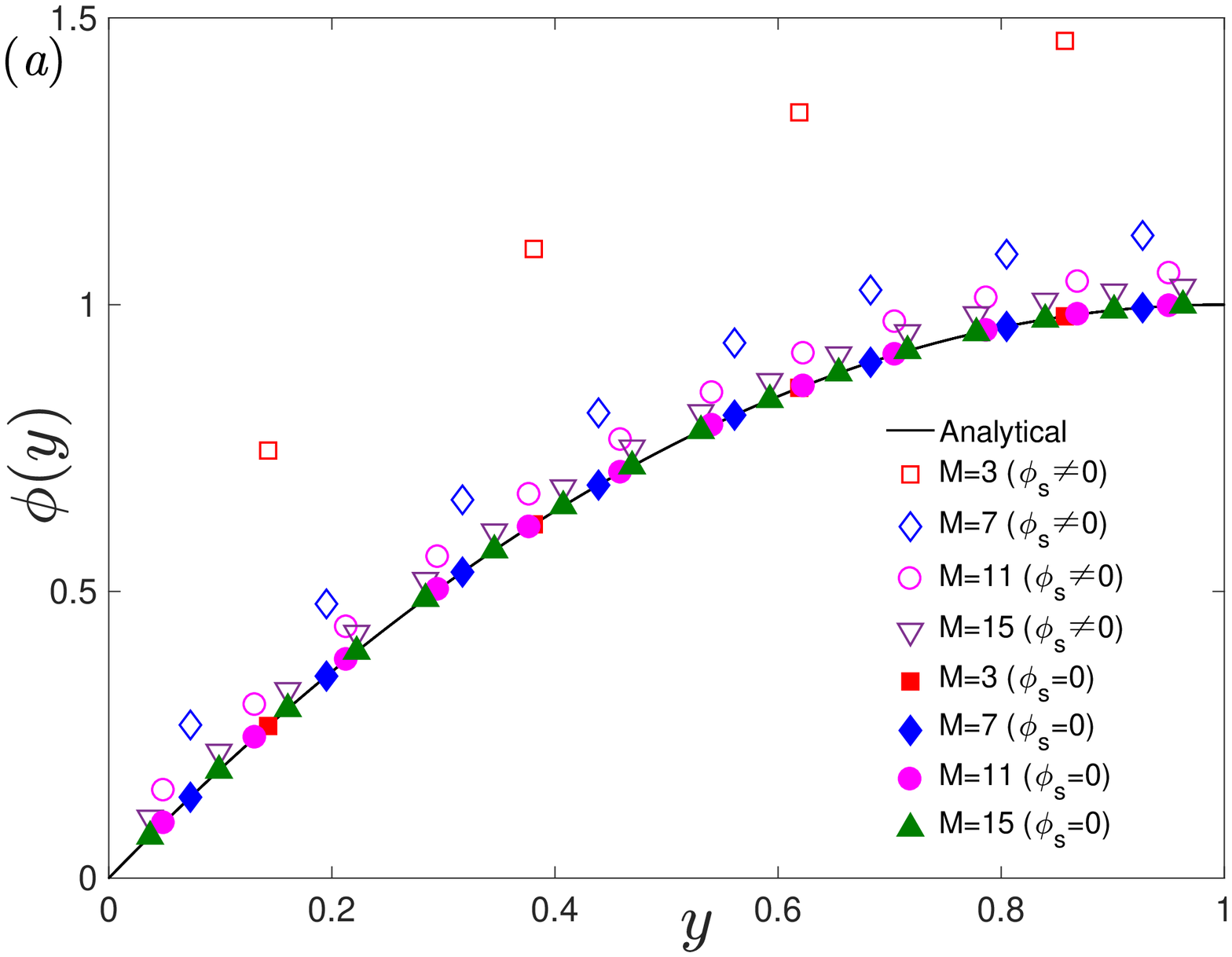}
\includegraphics[width=0.55\textwidth,height=0.3\textheight]{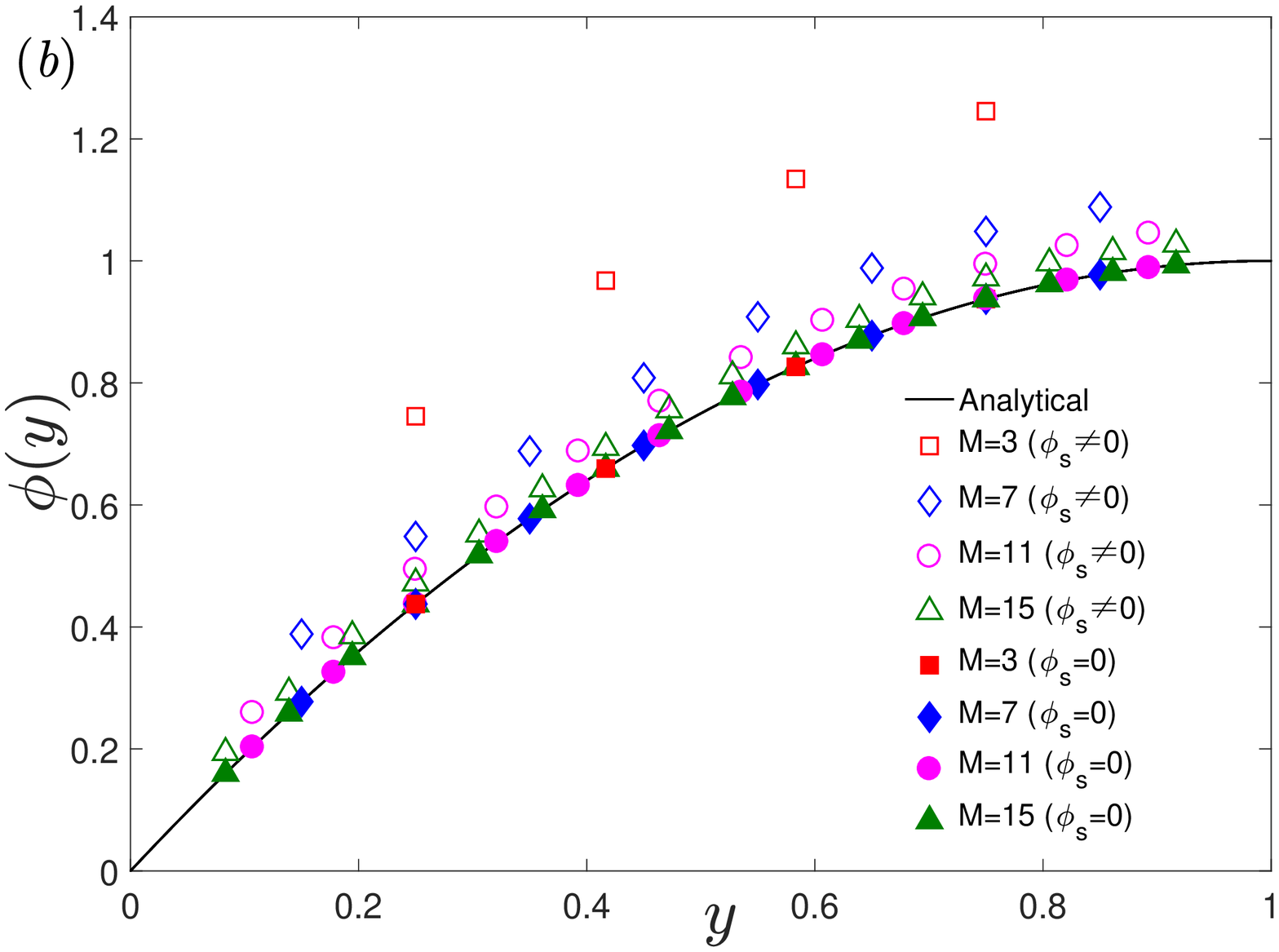}
%(a)&(b)\\
\end{tabular}
\caption{Profiles of scalar variable $\phi$ under different lattice sizes and [(a) $\gamma=0.6$; (b) $\gamma=1.5$] from the D2Q9 lattice model. Empty shapes denote the case that the relaxation time ($\tau_\phi=3.0$) dissatisfies $\phi_s=0$ with the distance ratio $\gamma$, while filled ones denote the case that $\tau_\phi$   satisfies $\phi_s=0$ with $\gamma$.}
\label{numSlipD2Q9}
\end{figure}

\begin{figure}
\begin{tabular}{cc}
\includegraphics[width=0.5\textwidth,height=0.3\textheight]{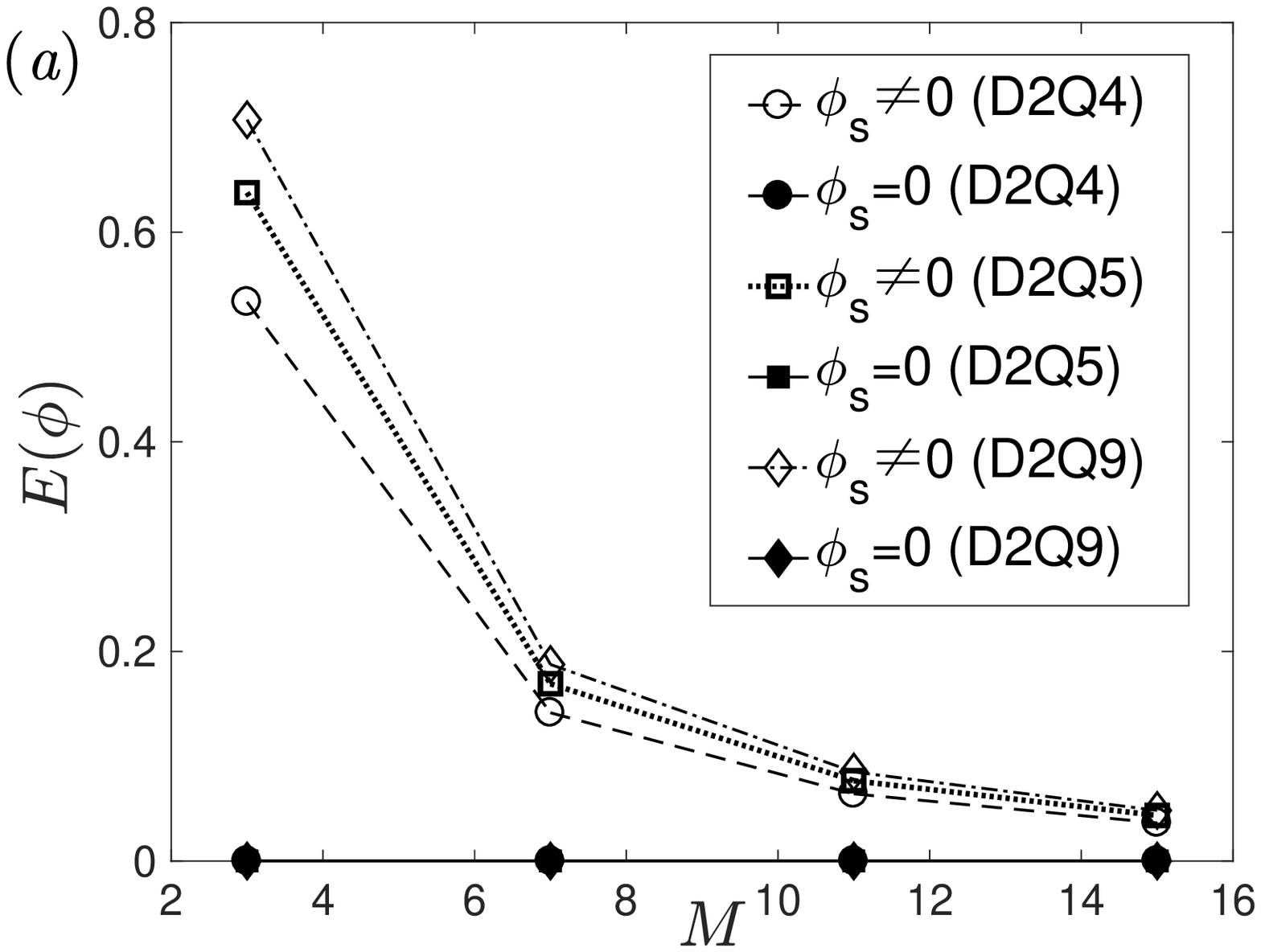}
\includegraphics[width=0.5\textwidth,height=0.3\textheight]{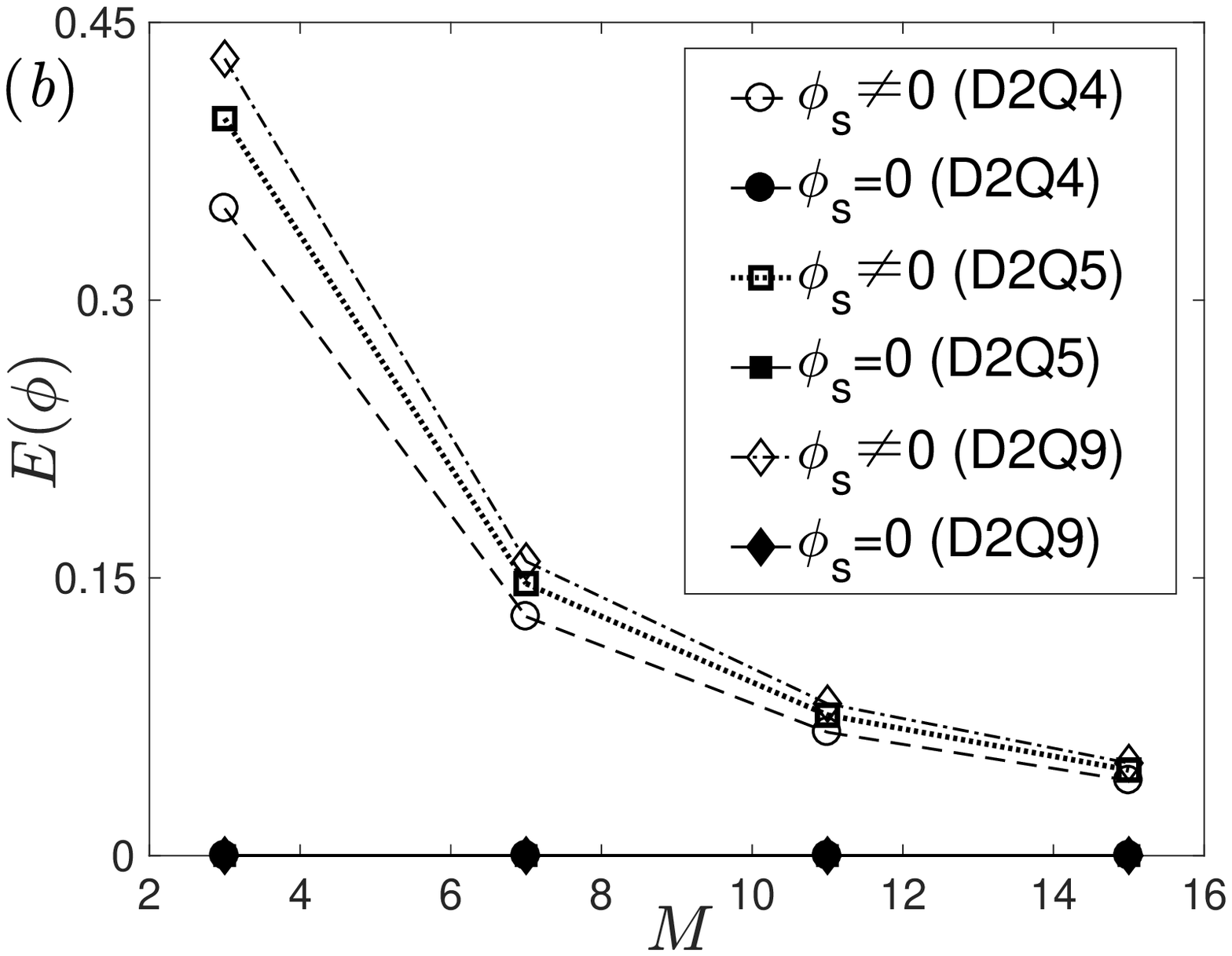}
%(a)&(b)\\
\end{tabular}
\caption{Relative error of $\phi$ against grid size $M$ and [(a) $\gamma=0.6$; (b) $\gamma=1.5$]. Dashed lines with empty shapes denote the case that the relaxation time ($\tau_\phi=3.0$) dissatisfies $\phi_s=0$ with $\gamma$, while solid lines with filled shapes denote the case that $\tau_\phi$   satisfies $\phi_s=0$ with $\gamma$.}
\label{RelaErr}
\end{figure}

The relations between $\gamma$ and $\tau_\phi$ are next examined especially for the numerical slip $\phi_s=0$. To this end, simulations with different grid sizes are carried out for two different values of $\tau_\phi$ at each of two distance ratios $\gamma=0.6$ and $\gamma=1.5$. One relaxation time is given by $\gamma$ to satisfy $\phi_s=0$ as derived above, while the other relaxation time (e.g., $\tau_\phi=3.0$) is not the case. For $\gamma=0.6$ and $\gamma=1.5$ considered here, the corresponding relaxation times $\tau_\phi$ to ensure $\phi_s=0$ can be obtained from Eq. \eqref{gamapsD2Q4} as $\tau_\phi=\frac{4+\sqrt{37}}{10}$ and $\tau_\phi=\frac{\sqrt{13}-1}{2}$ for the D2Q4 model, Eq. \eqref{gamapsD2Q5} as $\tau_\phi=\frac{25+\sqrt{1105}}{60}$ and $\tau_\phi=\frac{\sqrt{370}-4}{12}$ for the D2Q5 model, and  Eq. \eqref{gamapsD2Q9} as $\tau_\phi=\frac{19}{20}$ and $\tau_\phi=\frac{5}{4}$ for the D2Q9 model. Figs. \ref{numSlipD2Q4}, \ref{numSlipD2Q5} and \ref{numSlipD2Q9} respectively present the simulated results of the D2Q4, D2Q5 and D2Q9 lattice models.
As clearly shown in the figures, only when the relaxation time $\tau_\phi$ is determined by $\gamma$ while guaranteeing $\phi_s=0$, the results of the BGK model agree well with the analytical solution even with four grid points. However, if this requirement is not satisfied, clear discrepancies between the LBE results and the analytical solution can be observed even at $M=15$. Furthermore, to quantify the differences between such two results, the relative errors in $L^1$ norm are evaluated under different grid sizes. In Fig. \ref{RelaErr}, the relative errors $E(\phi)$ of the scalar variable $\phi$ are plotted against the grid size $M$.
It is clearly shown that as compared with the case of $\tau_\phi=3.0$, the obviously large errors are significantly reduced near zero when $\tau_\phi$ is given by $\gamma $ to ensure $\phi_s=0$. This further strengthens and supports our theoretical derivations.

The results exhibited in Fig. \ref{normalslip} has shown that the numerical slip of the NHABB boundary condition is different from that of the HABB boundary condition ($\gamma=1/2$). This indicates that the relaxation time $\tau_\phi$ derived from $\phi_s=0$ for $\gamma=1/2$ [Eqs. \eqref{gamapshalfD2Q4}, \eqref{gamapshalfD2Q5} and \eqref{gamapshalfD2Q9}] must be amended for the NHABB boundary condition to derive accurate results. In what follows, the values of relaxation time $\tau_\phi$ are inspected versus different values of $\gamma$ under the numerical slip $\phi_s=0$. In Tab. \ref{Tab:tauphi}, the approximations of the calculated values of $\tau_\phi$ from Eqs. \eqref{gamapsTD2Q4}, \eqref{gamapsTD2Q5} and \eqref{gamapsTD2Q9} are listed against $\gamma$.
\begin{table}[!hbp]
  \caption{Relaxation time $\tau_\phi$ versus distance ratio $\gamma$ for $\phi_s=0$ with the D2Q4, D2Q5 and D2Q9 lattice models.}
  \vspace{0.4em}
  \label{Tab:tauphi}
  \centering
  \begin{tabular*}
  {16cm}{@{\extracolsep{\fill}}cccc}
 \toprule[0.04em]\toprule[0.04em]
 \multirow{2}{*}{$\gamma$} &  \multicolumn{3}{c}{$\tau_\phi$}\\
 \cline{2-4}
 &    D2Q4&  D2Q5&  D2Q9 \\
 \midrule[0.04em]
 0.2&   $\frac{8+\sqrt{13}}{10}\approx 1.1606$&   $\frac{45+\sqrt{345}}{60}\approx 1.0596$& $\frac{29+\sqrt{129}}{40}\approx 1.0090$ \\
 0.5&   $1.0000$&    $\frac{6+\sqrt{30}}{12}\approx0.9564$&     $\frac{2+\sqrt{3}}{4}\approx0.9330$ \\
 0.8&   $\frac{2+\sqrt{73}}{10}\approx 1.0544$&    $\frac{15+\sqrt{2145}}{60}\approx1.0219$&  $\frac{11+\sqrt{849}}{40}\approx 1.0034$ \\
 1.2&   $\frac{-2+\sqrt{193}}{10}\approx 1.1892$&    $\frac{-5+\sqrt{5545}}{60}\approx 1.1578$&  $\frac{-1+3\sqrt{241}}{40}\approx1.1393$ \\
 1.5&   $\frac{\sqrt{13}-1}{2} \approx 1.3028$&    $\frac{\sqrt{370}-4}{12}\approx1.2696$&     $1.2500$ \\
 %1.8&   $\frac{\sqrt{13}-1}{2} \approx 1.3028$&    $\frac{\sqrt{370}-4}{12}\approx1.2696$&     $1.2500$ \\
 \bottomrule[0.04em]
 \bottomrule[0.04em]
\end{tabular*}
\end{table}
As seen from the table, the relaxation times $\tau_\phi$ are approximate to 1.0 at $\gamma=0.5$ in the D2Q5 and D2Q9 lattice model. Thus, as have pointed out in Ref. \cite{ShuC16}, satisfactory results can be usually obtained as $\tau_\phi=1.0$ even if the numerical slip is not strictly removed. However, we should note that this result is derived and valid for the halfway boundary scheme (i.e., $\gamma=0.5$). In fact, when the distance ratio $\gamma$ deviates away from 0.5, e.g., $\gamma=1.5$, the relaxation time $\tau_\phi$ is definitely larger than 1.0, which is also reflected in Fig. \ref{Depgatau}. This clearly indicates that when $\gamma$ varies away from 0.5, accurate results cannot be achieved any longer if $\tau_\phi$ still remains at 1.0. In other words, to derive accurate results ($\phi_s=0$), the relaxation time $\tau_\phi$ must be adjusted with $\gamma$.

\subsection{Diffusion between two concentric cylinders}
\begin{figure}
\centering
\includegraphics[width=0.55\textwidth]{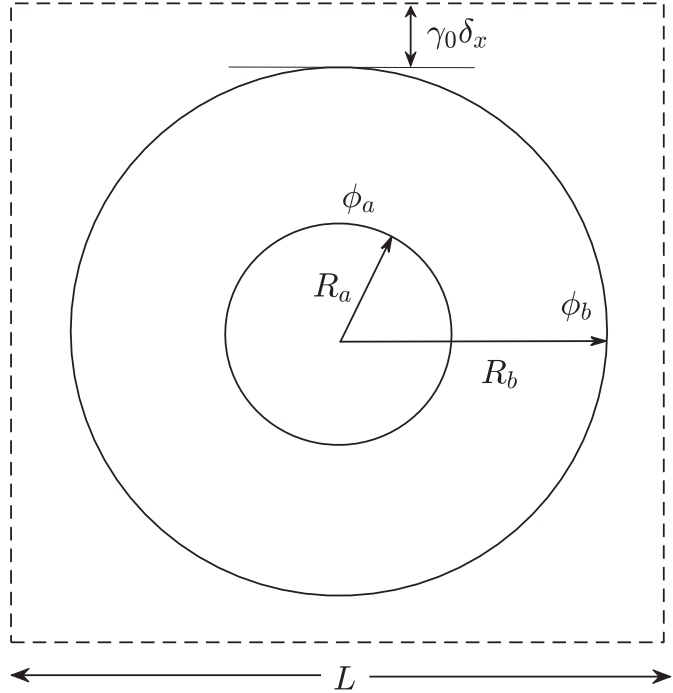}
\caption{Schematic of diffusions between two concentric cylinders.}
\label{ScheDiffCylinder}
\end{figure}

In this section, we investigate a more complex problem, i.e., the steady diffusion between two concentric circular cylinders, as shown in Fig. \ref{ScheDiffCylinder}.
The inner cylinder has radius of $R_a$ and boundary value $\phi_a$, while the outer cylinder has radius of $R_b$ and boundary value $\phi_b$. The outer cylinder boundary is separated from the square region with a distance of $\gamma_0\delta_x$. There is no source term for diffusions between the two concentric cylinders. From Eq. \eqref{CDES} in polar coordinates, the analytical solution to this problem can be solved and read as  \cite{{Carslaw13}}
\begin{equation}
  \phi(r)=\frac{\phi_a \text{ln}(R_b/r)+\phi_b \text{ln}(r/R_a)}{\text{ln}(R_b/R_a)},  \quad R_a\leq r \leq R_b.
\end{equation}

In the simulations, the two cylinders are positioned at the center of a square region with length $L=1.0$. The radius ratio of the two cylinders is $R_a:R_b=1:2$, the diffusion coefficient is set to $D=0.001$ and the boundary values are $\phi_a=0.0, \phi_b=1.0$. Unlike the previous problem with straight walls, the curved boundary geometries herein may bring different distance ratios, denoted by $\gamma_{in}$ and $\gamma_{out}$ for the boundary nodes respectively of the inner and outer cylinders. To have an unique relaxation time $\tau_\phi$ in ensuring $\phi_s=0$ (see Eqs. \eqref{gamapsD2Q4}, \eqref{gamapsD2Q5} and \eqref{gamapsD2Q9}), we approximate the distance ratio $\gamma$ by the average values of all $\gamma_{in}$ and $\gamma_{out}$ at a given $\gamma_0$ and grid number $M$.

\begin{figure}
\begin{tabular}{cc}
\includegraphics[width=0.55\textwidth,height=0.3\textheight]{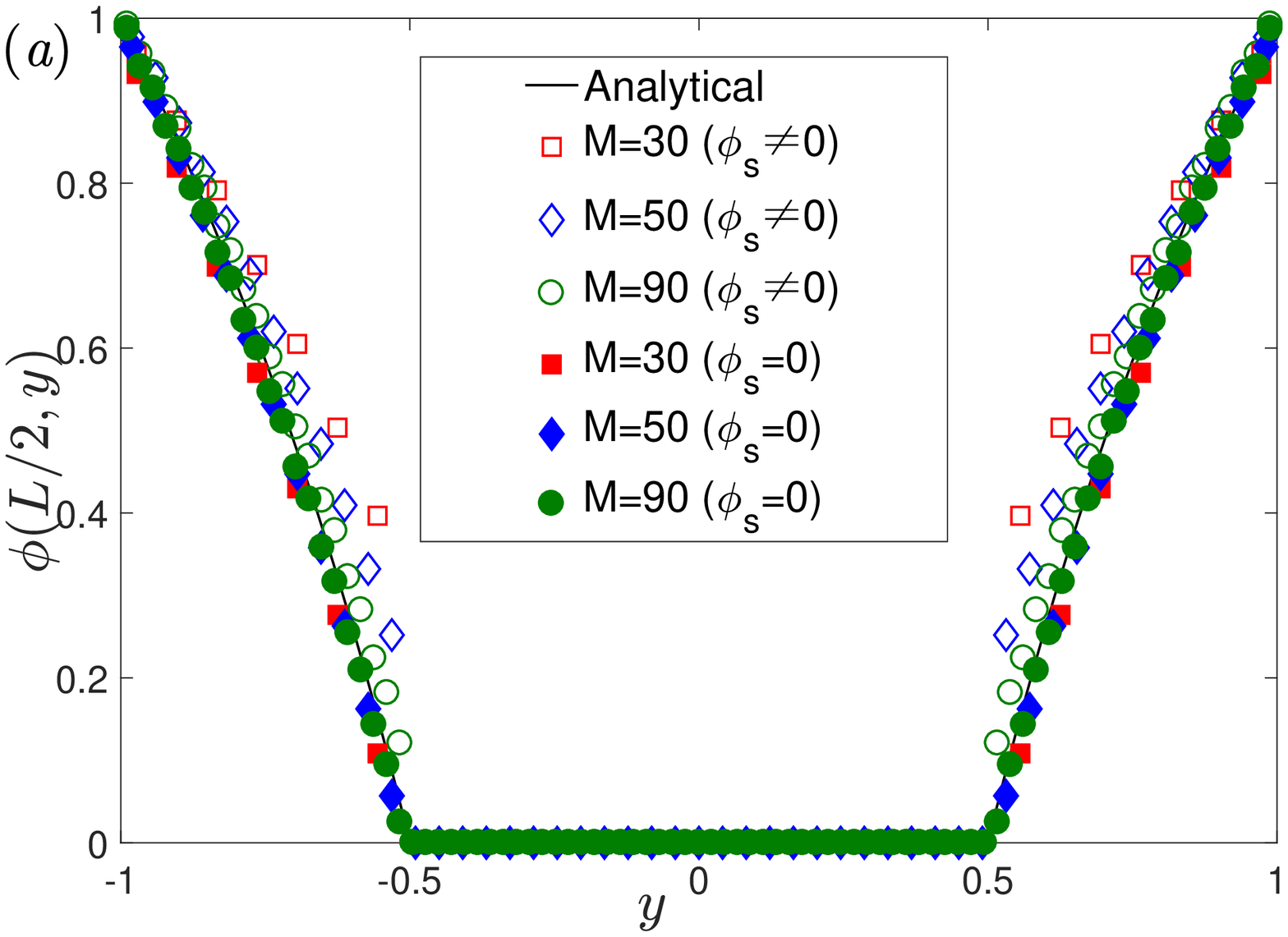}
\includegraphics[width=0.55\textwidth,height=0.3\textheight]{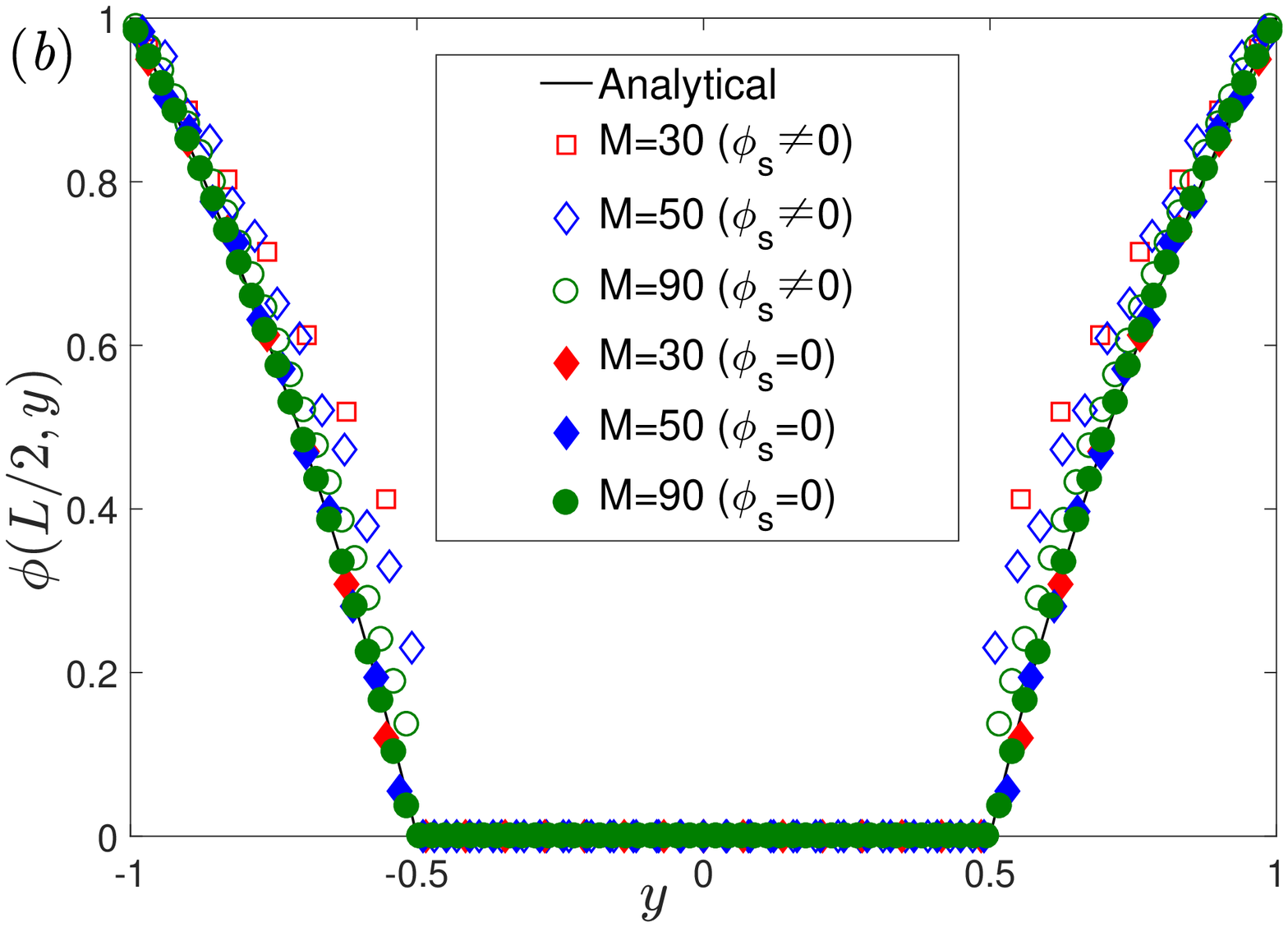}
%(a)&(b)\\
\end{tabular}
\caption{Profiles of scalar variable $\phi$ under different lattice sizes and [(a) $\gamma_0=0.2$; (b) $\gamma_0=0.8$] from the D2Q4 lattice model. Empty shapes denote the case that the relaxation time ($\tau_\phi=8.0$) dissatisfies $\phi_s=0$ with the average distance ratio $\gamma$, while filled ones denote the case that $\tau_\phi$ satisfies $\phi_s=0$ with $\gamma$.}
\label{cylinderD2Q4}
\end{figure}

\begin{figure}
\begin{tabular}{cc}
\includegraphics[width=0.55\textwidth,height=0.3\textheight]{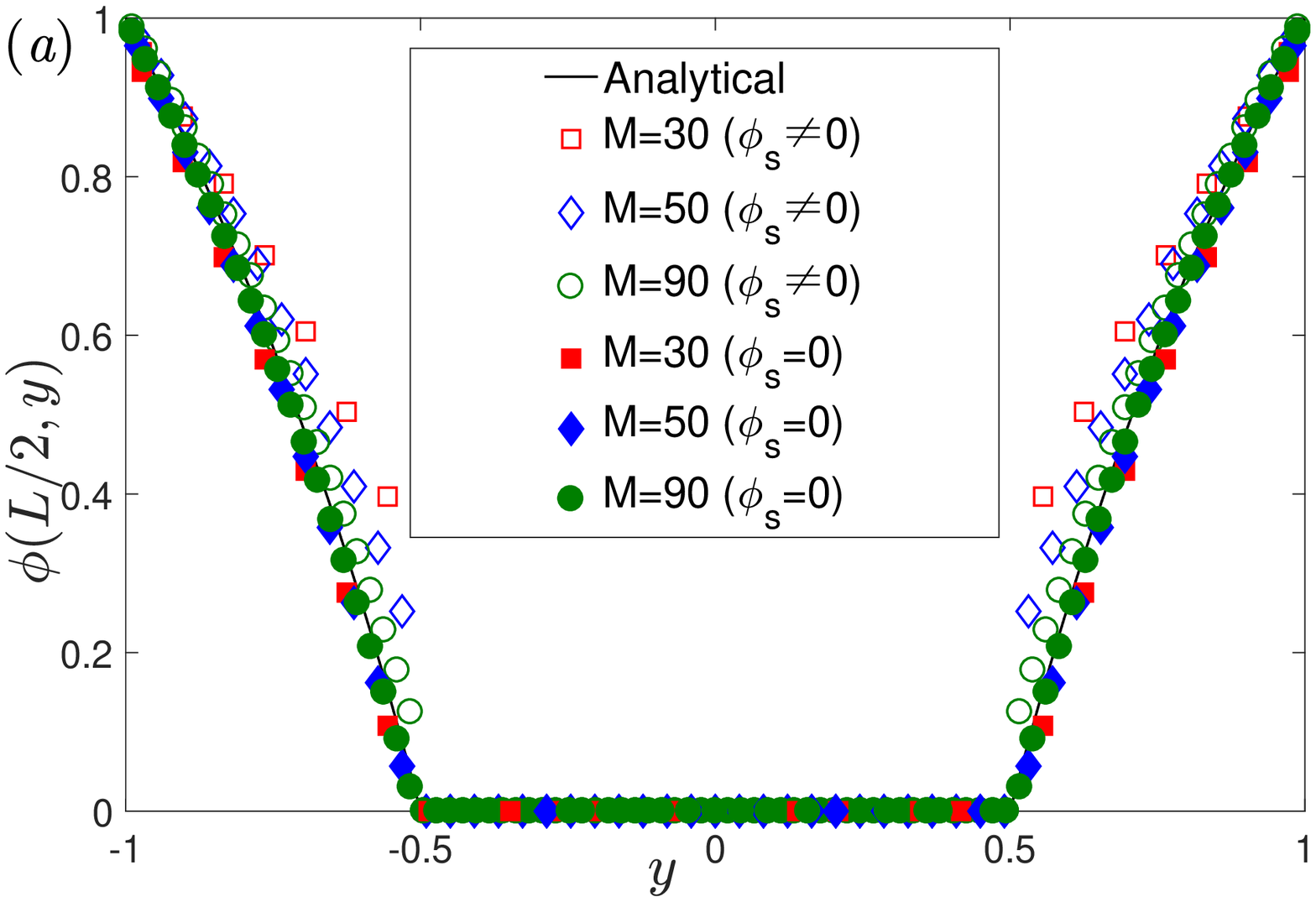}
\includegraphics[width=0.55\textwidth,height=0.3\textheight]{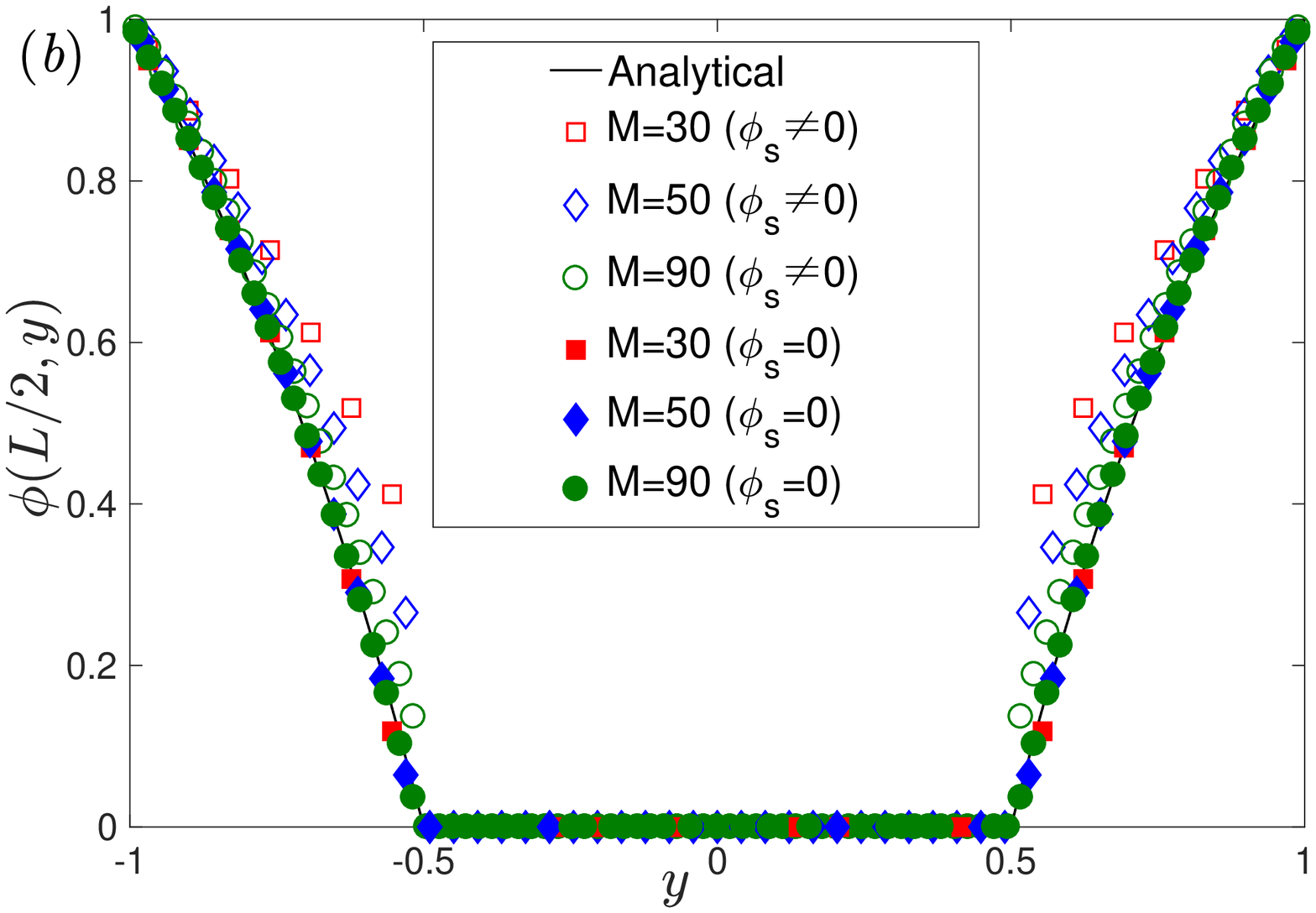}
%(a)&(b)\\
\end{tabular}
\caption{Profiles of scalar variable $\phi$ under different lattice sizes and [(a) $\gamma_0=0.2$; (b) $\gamma_0=0.8$] from the D2Q5 lattice model. Empty shapes denote the case that the relaxation time ($\tau_\phi=8.0$) dissatisfies $\phi_s=0$ with the average distance ratio $\gamma$, while filled ones denote the case that $\tau_\phi$ satisfies $\phi_s=0$ with $\gamma$.}
\label{cylinderD2Q5}
\end{figure}

\begin{figure}
\begin{tabular}{cc}
\includegraphics[width=0.55\textwidth,height=0.3\textheight]{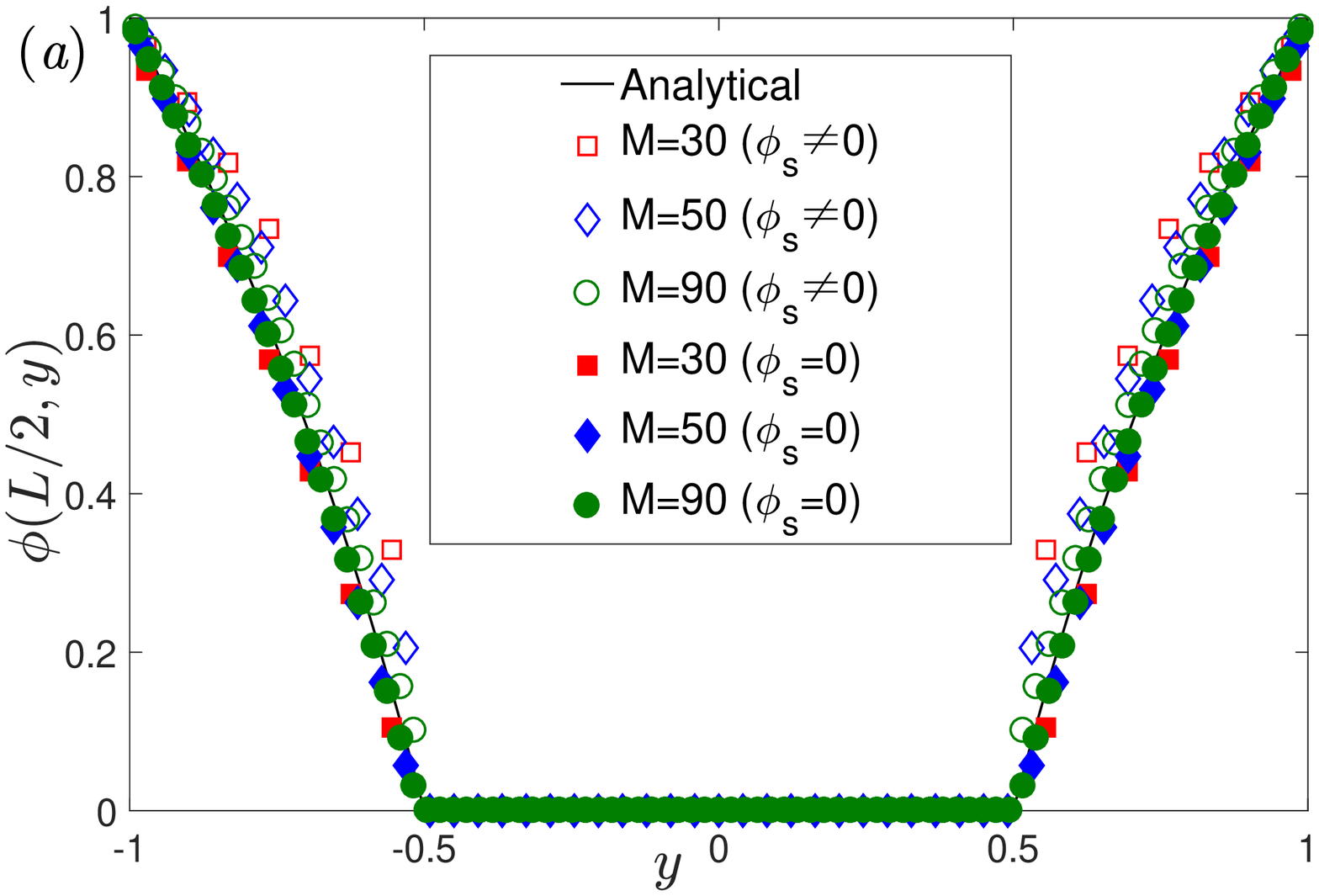}
\includegraphics[width=0.55\textwidth,height=0.3\textheight]{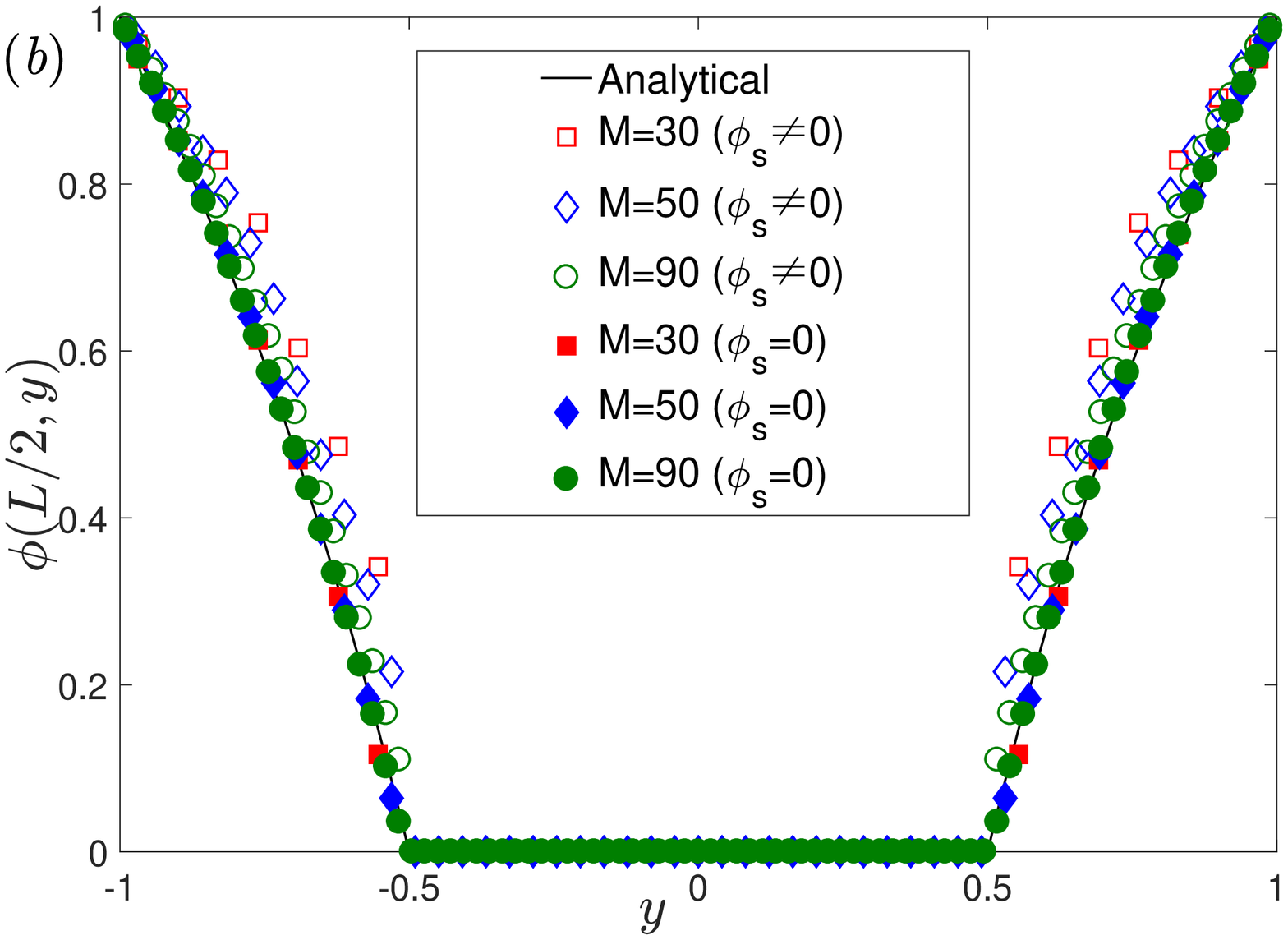}
%(a)&(b)\\
\end{tabular}
\caption{Profiles of scalar variable $\phi$ under different lattice sizes and [(a) $\gamma_0=0.2$; (b) $\gamma_0=0.8$] from the D2Q9 lattice model. Empty shapes denote the case that the relaxation time ($\tau_\phi=8.0$) dissatisfies $\phi_s=0$ with the average distance ratio $\gamma$, while filled ones denote the case that $\tau_\phi$ satisfies $\phi_s=0$ with $\gamma$.}
\label{cylinderD2Q9}
\end{figure}

Two cases of $\gamma_0$, i.e., $\gamma_0=0.2, 0.8$, are simulated with two relaxation times $\tau_\phi$. As done in the above problem, one $\tau_\phi$ is given by the average $\gamma$ from each $\gamma_0$ to meet $\phi_s=0$. The distributions of $\phi$ along the centerline are predicted by the D2Q4, D2Q5 and D2Q9 lattice models. Figs. \ref{cylinderD2Q4}, \ref{cylinderD2Q5} and \ref{cylinderD2Q9} delineate the profiles of $\phi$ between the two cylinders under different grid sizes $M$.
Take the D2Q4 lattice model as an example. When the relaxation time $\tau_\phi$ obeys Eq. \eqref{gamapsD2Q4} even through the average $\gamma$, the predictions are all much more accurate than the cases where $\tau_\phi$ dissatisfies $\phi_s=0$ (e.g., $\tau_\phi=8.0$). The similar results in Figs. \ref{cylinderD2Q5} and \ref{cylinderD2Q9} from the D2Q5 and D2Q9 lattice models again demonstrate our analysis in this work. In addition, with a careful look at Fig. \ref{numSlipD2Q4} and Fig. \ref{cylinderD2Q4} for $\phi_s=0$ under a small grid number, the agreement of the present results with the analytical solution is found not as closely as those achieved in the previous problem. This result can be expected since the approximated average distance ratio is used for the curved boundary of the cylinders.

\section{Conclusions} \label{results}
In this work, the discrete effect on the ABB boundary condition has been analyzed in the framework of BGK model for the CDE. Different from previous works on the HABB boundary condition, the boundary scheme adopted in this paper incorporates the distance ratio $\gamma$ of boundary nodes as a free parameter \cite{HuangJ15}. The theoretical derivations clearly shows that unlike the HABB boundary scheme ($\gamma=1/2$), the numerical slip $\phi_s$ of the NHABB boundary condition can be relieved from only relating with the relaxation time $\tau_\phi$ but together with $\gamma$. Therefore, as the numerical slip $\phi_s=0$ is guaranteed, the relaxation time $\tau_\phi$ can be freely adjusted as a function of the distance ratio $\gamma$, which cannot be realized for the HABB boundary condition. Concretely, for the distance ratio $\gamma$ varying in a proper range, if the relaxation time $\tau_\phi$ changing with $\gamma$ conforms to Eq. \eqref{gamapsD2Q4} in the D2Q4 lattice model, Eq. \eqref{gamapsD2Q5} in the D2Q5 lattice model, or Eq. \eqref{gamapsD2Q9} in the D2Q9 lattice model, the discrete effect of the NHABB boundary condition can be eliminated within the framework of BGK model, while in the HABB boundary condition, the discrete effect always exists except for a special value of the relaxation time $\tau_\phi$. On the basis of the BGK model, the non-halfway and halfway ABB boundary conditions are both implemented to validate the theoretical analysis. For the unidirectional diffusion with a parabolic distribution in a straight channel, the numerical results show that owing to the free parameter of $\gamma$, a much wider range of the relaxation time $\tau_\phi$ can be achieved to produce accurate results. For the diffusion between two concentric circular cylinders, satisfactory agreements between the numerical results and the analytical solution can be obtained even with the average distance ratio.

We would like to point out that due to the quadratic dependence on $\gamma$ in $\phi_s=0$, the minimum relaxation time $\tau_\phi$ can reach only around 1 while not near 0.5, as shown in Fig. \ref{Depgatau}. However, we also note that this limitation can be improved by adding more degree of freedom in determining $\tau_\phi$ from the numerical slip $\phi_s$. One straightforward strategy for this is to extend the present analysis from the framework of BGK model to the two-relaxation-time (TRT) or the multiple-relaxation-time (MRT) model. This topic will be investigated in our forthcoming work.
 \vspace{5mm}
\begin{acknowledgments}
This work is financially supported by the National Natural Science Foundation of China (No. 51776068, No. 51606064 and No. 11602075) and the Fundamental Research Funds for the Central Universities (No. 2018MS060). L. Wang would like to thank Profs. Wen-An Yong, Zhaoli Guo and Dr. Weifeng Zhao for their fruitful discussions and advices.
\end{acknowledgments}
%

%% The \nocite command causes all entries in a bibliography to be printed out
%% whether or not they are actually referenced in the text. This is appropriate
%% for the sample file to show the different styles of references, but authors
%% most likely will not want to use it.
\nocite{*}


\end{document}